\documentclass[
 reprint,
 groupedaddress,
 amsmath,amssymb,
 aps,
 prd,
 longbibliography,
 noeprint
]{revtex4-2}

\usepackage{hyperref}
\usepackage{soul}

\usepackage{graphicx}
\usepackage{dcolumn}
\usepackage{bm}
\usepackage[dvipsnames]{xcolor}

\newcommand{\Lya}{Ly$\alpha$}
\newcommand{\HI}{$\textrm{HI}$}

\newcommand{\orcidicon}[1]{\href{https://orcid.org/#1}{\includegraphics[height=\fontcharht\font`\B]{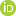}}}

\DeclareMathOperator\erf{erf}

\begin{document}

\title{Dynamical Dark Energy Imprints in the Lyman-$\alpha$ Forest}

\author{Diego Garza\, \orcidicon{0009-0003-0226-6988}}
\email{Contact author: digarza@ucsc.edu}
\affiliation{Department of Astronomy and Astrophysics, University of California, Santa Cruz, 1156 High Street, Santa Cruz, CA 95064}

\author{Brant Robertson\, \orcidicon{0000-0002-4271-0364}}
\affiliation{Department of Astronomy and Astrophysics, University of California, Santa Cruz, 1156 High Street, Santa Cruz, CA 95064}

\author{Piero Madau\, \orcidicon{0000-0002-6336-3293}}
\affiliation{Department of Astronomy and Astrophysics, University of California, Santa Cruz, 1156 High Street, Santa Cruz, CA 95064}
\affiliation{Dipartimento di Fisica ``G. Occhialini," Università degli Studi di Milano-Bicocca, Piazza della Scienza 3, I-20126 Milano, Italy}

\author{Nick Gnedin\, \orcidicon{0000-0001-5925-4580}}
\affiliation{Department of Astronomy and Astrophysics, University of Chicago, 5640 South Ellis Avenue, Chicago, IL 60637}
\affiliation{Theory Division, Fermi National Accelerator Laboratory, PO Box 500, MS 106, Batavia, IL 60510-5011}
\affiliation{Kavli Institute for Cosmological Physics, University of Chicago, 5640 South Ellis Avenue, Chicago, IL 60637}

\author{Matthew W. Abruzzo\, \orcidicon{0000-0002-7918-3086}}
\affiliation{Department of Physics and Astronomy, University of Pittsburgh, 3941 O'Hara St, Pittsburgh, PA 15260}

\author{Reuben D. Budiardja\, \orcidicon{0000-0003-0395-8532}}
\affiliation{National Center for Computational Sciences, Oak Ridge National Laboratory,
1 Bethel Valley Road, Oak Ridge, TN 37830}

\author{Robert Caddy\, \orcidicon{0000-0002-4475-3181}}
\affiliation{Department of Astrophysical Sciences, Princeton University, Peyton Hall, 4 Ivy Lane,  Princeton, NJ 08544}

\author{Evan Schneider\, \orcidicon{0000-0001-9735-7484}}
\affiliation{Department of Physics and Astronomy, University of Pittsburgh, 3941 O'Hara St, Pittsburgh, PA 15260}

\author{James~B.~White~III\, \orcidicon{0009-0005-2186-075X}}
\affiliation{National Center for Computational Sciences, Oak Ridge National Laboratory,
1 Bethel Valley Road, Oak Ridge, TN 37830}

\author{Bruno Villasenor\, \orcidicon{0000-0002-7460-8129}}
\affiliation{Advanced Micro Devices, 2485 Augustine Drive, Santa Clara, California 95054}

\date{\today}

\begin{abstract}
   The nature of dark energy (DE) remains elusive, even though it constitutes the dominant energy-density component of the Universe and drives the late-time acceleration of cosmic expansion. By combining measurements of the expansion history from baryon acoustic oscillations, supernova surveys, and cosmic microwave background data, the Dark Energy Spectroscopic Instrument (DESI) Collaboration has inferred that the DE equation of state may evolve over time. The profound implications of a time-variable, ``dynamical" DE (DDE) that departs from a cosmological constant motivate the need for independent observational tests. In this work, we use cosmological hydrodynamical simulations of structure formation to investigate how DDE affects the properties of the Lyman-$\alpha$ ``forest'' of absorption features produced by neutral hydrogen in the cosmic web. We find that DDE models consistent with the DESI constraints induce a spectral tilt in the forest transmitted flux power spectrum, imprinting a scale- and redshift-dependent signature relative to standard $\Lambda$CDM cosmologies. These models also yield higher intergalactic medium temperatures and reduced Lyman-$\alpha$ opacity compared to $\Lambda$CDM. We discuss the observational implications of these trends as potential avenues for independent confirmation of DDE.
\end{abstract}
\maketitle

\section{Introduction}

    Over the past 25 years, the spatially flat $\Lambda$ cold dark matter ($\Lambda$CDM) model has become the dominant cosmological paradigm, as it provides the most consistent fit to a wide range of observations. In this model, the formation and evolution of large-scale structure are governed by the perturbation theory of general relativity \citep{2002PhR...367....1B} and seeded by primordial inhomogeneities in a cold, collisionless dark matter component. This dark matter accounts for approximately 25\% of the Universe’s total energy-density budget and clusters gravitationally to form the scaffolding of cosmic structure.

In contrast, dark energy (DE) is a smooth, non-clustering component that currently dominates the energy density of the Universe. Dark energy accounts for the observed late-time acceleration of cosmic expansion, as first revealed by Type Ia supernovae \citep{DEDiscoveryReiss, DEDiscoveryPerlmutter}. While the gravitational effects of dark matter are well understood, the fundamental physics of DE remain poorly constrained. The impact of DE on the cosmic expansion can be described by an equation of state parameter $w = P/\rho$, which relates its pressure to its energy density. In the $\Lambda$CDM framework, DE is modeled as a cosmological constant with $w = -1$, representing the energy density of the vacuum.

To allow for possible departures from this simple scenario, phenomenological extensions consider a time-varying equation of state, typically parameterized by $w$ and its derivative with respect to the scale factor, ${\rm d}w/{\rm d}a$. Constraining these parameters requires precise measurements of the expansion history across cosmic time. Observations of the cosmic microwave background (CMB) anisotropies \citep{Planck-Cosmo}, baryonic acoustic oscillations (BAO) \citep{BAO-one, eBOSS, DESI-DR2-Cosmo}, and Type Ia supernovae \citep{2022ApJ...933...59V, 2025ApJ...986..231R, 2024ApJ...962..113M} provide complementary constraints at different redshifts. Together with other cosmological probes, these datasets enable stringent tests of $\Lambda$CDM and offer sensitivity to deviations from the standard expansion history that may signal new fundamental physics or modifications to general relativity \citep{DETF, DEProbes}.
    
   With its first data release comprising over 18.7 million spectroscopic redshifts, the Dark Energy Spectroscopic Instrument (DESI) Collaboration has already advanced precision cosmology to a new frontier \citep{DESI-DR1-proper}. The inference of cosmological parameters from the first data release (DR1), based on baryon acoustic oscillation (BAO) measurements and their combination with big-bang nucleosynthesis constraints \citep{BBN}, cosmic microwave background data \citep{Planck-Cosmo}, and Type Ia supernova samples \citep{2022ApJ...938..110B, 2025ApJ...986..231R, 2024ApJ...973L..14D}, provides indications of a time-dependent DE equation of state \citep{DESI-DR1-Cosmo}. In its second data release (DR2), which nearly doubles the number of galaxies and quasars used, the DESI Collaboration strengthens the evidence for a dynamical dark energy (DDE) scenario \citep{DESI-DR2-Cosmo}, reporting at least a 2.8$\sigma$ deviation from the $\Lambda$CDM model when the DE equation of state is modeled as a linear function of the scale factor.
        
   Studies of the expansion history and the growth of structure are natural avenues for seeking additional verification of a time-evolving DE. The diffuse nature of the intergalactic medium (IGM), comprised of the baryonic matter residing between virialized dark matter halos, offers a sensitive environment for investigating the signatures of a dynamical DE component. The thermal history of the IGM is governed primarily by adiabatic cooling from cosmic expansion and by photoheating from hydrogen and helium ionization \citep{rad-cooling}. A key observable for probing the physical state of the IGM is the absorption of light from distant quasars by intervening patches of neutral hydrogen (\HI) along the line of sight (LOS), producing rest-frame Lyman-$\alpha$ (\Lya) absorption features that result in a characteristic ``forest’’ of lines redward of the observed \Lya\ transition \citep{1965ApJ...142.1633G, 1992MNRAS.255..319W}. The principal statistic adopted in this study to characterize the structure of the \Lya\ forest and constrain cosmological parameters is the LOS-averaged transmitted flux power spectrum (FPS), which provides high-resolution measurements down to comoving scales of $\approx 1\,h^{-1}\,\mathrm{Mpc}$ \citep{Croft_1998, 2005ApJ...635..761M}.
    
   In light of the DESI inference of DDE, we investigate several observationally motivated DDE cosmological models using high-resolution hydrodynamic simulations to identify potential signatures that distinguish them from $\Lambda$CDM. The effects of differing expansion histories are expected to manifest most clearly in moderate and underdense regions of the cosmic web, where the balance between adiabatic cooling and photoheating from the cosmic UV background dominates over virial heating. To assess possible imprints of DDE in these environments, we extract $\mathcal{O}(10^6)$ synthetic \Lya\ absorption spectra across several redshifts, enabling a detailed study of the scale- and redshift-dependent structure of the IGM. We compute relative differences in the IGM thermal state and the FPS to quantify the impact of the three DDE models reported by DESI DR1.

    Two prior studies \citep{Viel2003, Coughlin} have investigated the effects of a non-cosmological constant dark energy equation of state ($w \neq -1$) on the IGM, the forest, and the FPS. Both studies precede the DESI results suggesting a dynamical dark energy component. In \citep{Viel2003}, the optical depth distribution of the IGM and the resulting FPS are analyzed for cosmologies with a constant equation of state parameter $w \neq -1$. Using semi-analytical methods, the authors find that variations in the Hubble parameter $H(z)$ and the linear growth factor lead to a reduction in the \Lya\ optical depth, while largely preserving the shape of the FPS. For a model with $w = -0.4$, they report an enhancement of the FPS amplitude relative to $\Lambda$CDM by approximately an order of magnitude. They argue that this effect would be observable, provided one has an independent and precise characterization of the IGM’s thermal state, ionization conditions, and baryon fraction to accurately constrain $H(z)$.

    The authors of \cite{Coughlin} build upon this earlier work by performing high-resolution dark-matter-only $N$-body simulations to investigate the impact of DDE (${\rm d}w/{\rm d}a \neq 0$) on the \Lya\ FPS. To model baryonic effects, \cite{Coughlin} apply temperature prescriptions to ``field" and ``halo" particles, making use of a power-law relation between temperature and density in the IGM and the bulk virial properties of dark matter halos. They explore the boundaries of the parameter space allowed by the Planck constraints on DDE models and find only a very small impact on the \Lya\ FPS, as assessed using a $k$-sample Anderson–Darling test.

Here, we present an initial study of the IGM to investigate potential imprints of DDE on baryonic features of the Universe, including the mean IGM temperature, the effective \Lya\ optical depth, and the FPS. In Section~\ref{sec:meth}, we describe the hydrodynamic simulations used in this work, detailing the adopted cosmological parameters, the implementation of the DDE expansion history, and the methodology for measuring the effective optical depth. In Section~\ref{sec:res}, we present our results on the structure of the \Lya\ forest and quantify the impact of DDE. In Section~\ref{sec:disc}, we offer a brief discussion of the implications of DDE for the evolution of baryonic cosmic structures.

\section{Methodology}\label{sec:meth}

   To investigate potential effects of DDE on baryonic structure formation, we model the relevant physics of the IGM using hydrodynamic cosmological simulations. In Section~\ref{subsec:sims-const}, we describe the simulation code and the parameter values shared across all runs. Section~\ref{subsec:exp-hist} outlines the influence of DDE on the expansion history and the linear growth factor. In Section~\ref{subsec:sims-var}, we detail the implementation of DDE cosmologies within our simulation framework. The method used to compute the \Lya\ optical depth in each simulation cell is presented in Section~\ref{subsec:opt-depth}.

    \begin{table*}[t]
            \caption{\label{tab:table1} Cosmological parameters used for our study, adopted from Table 3 of \cite{DESI-DR1-Cosmo}. From left to right, we describe the present-day matter energy-density $\Omega_{m,0}$, present-day DDE equation of state parameter $w_0$, the linear interpolation of the DDE equation of state to the early Universe $w_a$, present-day baryon energy-density $\Omega_{b,0}$, and present-day Hubble parameter $H_0$. We assume a spatially flat cosmology for all models. Our DESI+CMB+$\Lambda$CDM model is the flat $\Lambda \textrm{CDM}$ DESI+CMB cosmology reported in \cite{DESI-DR1-Cosmo}, where the ``CMB" indicates the inclusion of temperature and polarization data from \cite{Planck-Cosmo} and lensing data from \cite{Planck-Lens} and \cite{ACT-Lens}. We explore $w_0 w_a \textrm{CDM}$ cosmology alternatives which combine supernova datasets with CMB and DESI data. From \cite{Planck-Cosmo} Table 2, we also keep the baryonic energy-density result $\Omega_b h^2 = 0.02242$ for each cosmology as well as the matter power spectrum normalization $\sigma_8 = 0.8102$. We adopt the spectral index value of $n_s=0.9665$. To isolate the impact of a DDE model, we run reference simulations with $w = -1$ ($\Lambda$CDM) using the same cosmological parameters as in the corresponding DDE cases. These parameters are listed in the bottom three rows.}
            \begin{ruledtabular}
            \begin{tabular}{ccccccc}
            \textrm{Cosmology Name}& $\Omega_{m,0}$ & $w_0$ & $w_a$ & $\Omega_{b,0}$ & $H_0$ $\textrm{[km s}^{-1}$ $\textrm{Mpc}^{-1}$ \textrm{]}\\
            \colrule
            DESI+CMB+$\Lambda$CDM & 0.3069 & -1.0 & 0.0 & 0.04853 & 67.97 \\
            \hline
            DESI + CMB + Pantheon & 0.3085 & -0.827 & -0.75 & 0.04844 & 68.03 \\
            DESI + CMB + DESY5 & 0.3160 & -0.727 & -1.05 & 0.04958 & 67.24 \\
            DESI + CMB + Union3 & 0.3230 & -0.65 & -1.27 & 0.05065 & 66.53 \\
            \hline
            DESI + CMB + $\Lambda$Pantheon & 0.3085 & -1.00 & 0. & 0.04844 & 68.03 \\
            DESI + CMB + $\Lambda$DESY5 & 0.3160 & -1.00 & 0. & 0.04958 & 67.24 \\
            DESI + CMB + $\Lambda$Union3 & 0.3230 & -1.00 & 0. & 0.05065 & 66.53 \\
            \end{tabular}
            \end{ruledtabular}
        \end{table*}
        
    \subsection{Simulations}\label{subsec:sims-const}
        We perform the simulations for this study using the cosmological version of the \texttt{Cholla} code \citep{2015ApJS..217...24S, 2021ApJ...912..138V} on the Frontier supercomputer at the Oak Ridge Leadership Computing Facility \cite{Frontier-proceeding}. The implementation of DDE is described in Section~\ref{subsec:exp-hist}. \texttt{Cholla} is a GPU-native, massively parallel, grid-based hydrodynamic simulation code \citep{2015ApJS..217...24S}. The Riemann problem is solved in \texttt{Cholla} using Godunov-based solvers \citep{godunov:hal-01620642} with third-order spatial reconstruction \citep{WOODWARD1984115}. Its uniform grid structure provides a precise platform for studying the IGM, as it enables accurate resolution of diffuse regions of the Universe, in contrast with mass-based or adaptive refinement approaches that concentrate resolution in the densest environments. All physics, including the non-equilibrium chemical network for hydrogen, helium, and electron species, is computed on the GPU. Cooling and heating rates are adopted from the \texttt{GRACKLE} library \citep{GRACKLE}.
        
        We fix the baryonic energy density and the normalization of the matter power spectrum to the values inferred by \citep{Planck-Cosmo}, and adopt the corresponding primordial helium mass fraction, $Y_{\textrm{He}} = 0.2454$. Assuming a metal-free composition, the hydrogen mass fraction is then $Y_{\textrm{H}} = 0.7546$. Initial conditions are generated with the \texttt{MUSIC} code \citep{2011MNRAS.415.2101H}, using the transfer function from \citep{EiseHuTransfer}.  Simulations are evolved from redshift $z = 100$ to $z = 1$ on a uniform Cartesian grid with $N = 2048^3$ cells in a comoving box of volume $V = (50\ h^{-1}\ \textrm{Mpc})^3$, corresponding to a spatial resolution of $\Delta x = 24.4\ h^{-1}\ \textrm{kpc}$. For a resolution study of the \Lya\ forest and IGM thermodynamic properties using the \texttt{Cholla} code, see Appendix A of \citep{2021ApJ...912..138V} or \citep{2022ApJ...933...59V}. We evolve $N_p = 2048^3$ dark matter particles, each with mass $m_{\textrm{DM}} = \rho_{c,0} \, \Omega_{\textrm{DM},0} \, V / N_p$, which corresponds to $m_{\textrm{DM}} / (10^6\ h^{-1}\ M_\odot) \approx 1.0435$--$1.1000$, depending on the cosmology.
        
        We adopt the UV background field from \cite{2022ApJ...933...59V}, who fit the one dimensional FPS measured by eBOSS \citep{Chabanier_2019}, Keck \citep{2019ApJ...872..101B}, and VLT \citep{2019ApJ...872..101B, XQ-100-FPS}, finding that hydrogen and helium reionization is completed by $z \approx 6.0$ and $z \approx 3.0$ respectively. We output full snapshots at redshifts $z = [9, 6, 4, 3.6, 3.4, 3.2, 3, 2.8, 2.6, 2.4, 2.33, 2.2, 1.491, 1.317]$. We choose the $\Delta z=0.2$ redshift spacing between $z=3.6$ and $z=2.2$ to match the output redshifts from the DESI DR1 FPS measurement \citep{DESI-DR1-FPS}. We also save snapshots at the effective redshifts of the DESI DR1 tracers (from Table 1 of \cite{DESI-DR1-Cosmo} including emission line galaxies at $z_{\textrm{eff}}=1.317$, quasar spectra at $z_{\textrm{eff}}=1.491$, and \Lya\ quasar spectra at $z_{\textrm{eff}}=2.33$). Each simulation runs on 64 Frontier nodes, each with 8 GPUs, for a total of roughly 2.5 wall clock hours per simulation, equal to about 160 Frontier node-hours per simulation, excluding I/O overheads, and ending at redshift $z=1.3$. The entire set of simulations used in this study were completed in about 1120 Frontier node-hours.

        \begin{figure*}
            \centering
            \includegraphics[scale=0.525]{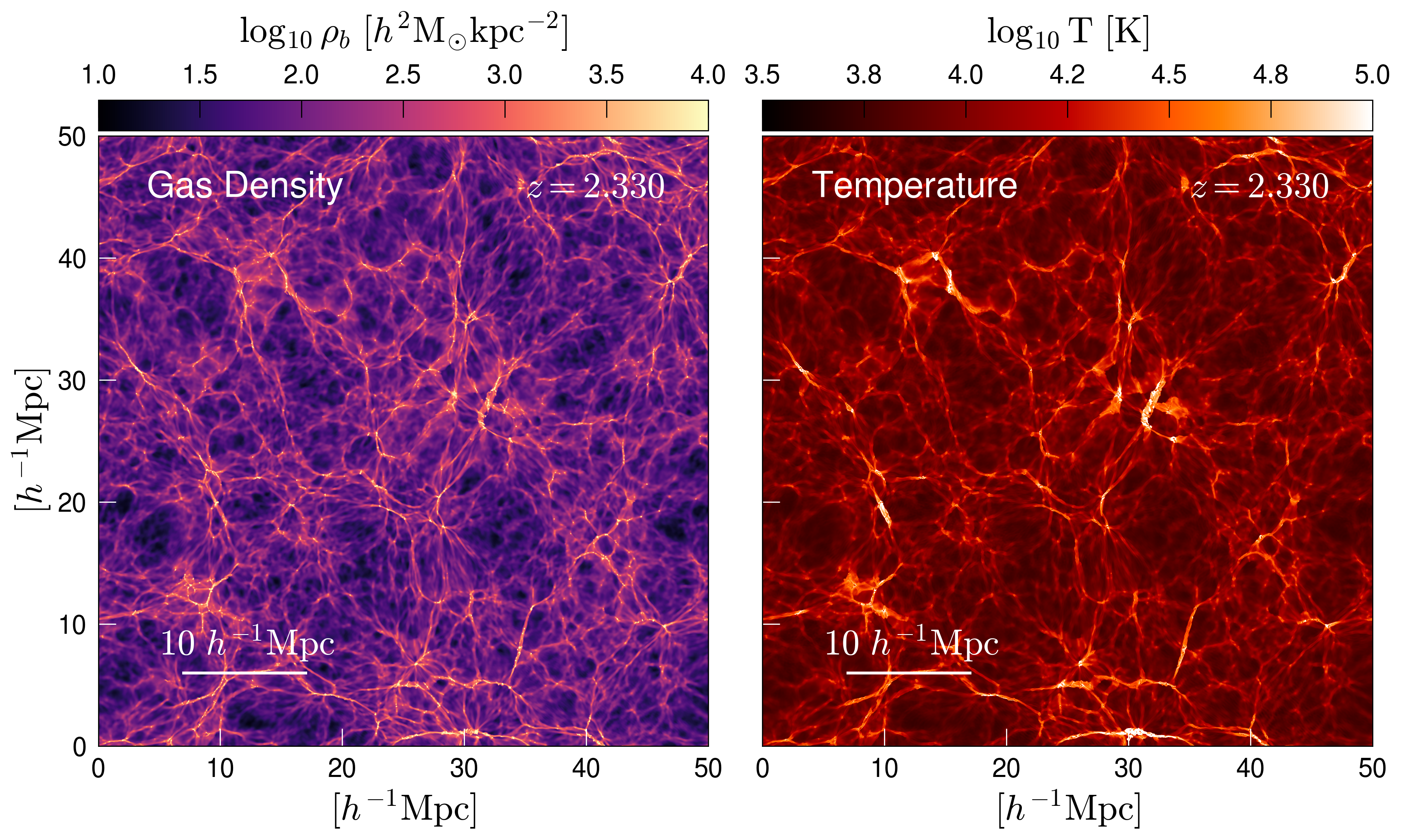}
            \caption{A two-dimensional slice of the DESI+CMB+$\Lambda$CDM cosmology at $z=2.33$ extending across the entire simulation box. The left panel shows the gas density projected along $500\ h^{-1}\textrm{kpc}$. The right panel shows the density-weighted temperature of the gas. The filamentary nature of large-scale structure is shown with the densest regions hosting the warmest portions of the projected gas distribution.}
            \label{fig:slab-viz}
        \end{figure*}                
        
    \subsection{Expansion History}\label{subsec:exp-hist}
        In a homogeneous and isotropic Universe, the expansion rate is determined by the Hubble parameter $H(t)$, which is governed by the Friedmann equation. To describe the expansion history of a given cosmological model, we compute the dimensionless age of the Universe, $H_0 t(z)$, as a function of redshift by solving

\begin{equation} \label{eqn:age}
    H_0 t(z) = \int_{z}^{\infty} \frac{\textrm{d}z'}{(1 + z') \, \xi(z')} \,,
\end{equation}

 \noindent where $H_0$ is the present-day value of the Hubble parameter and 

        \begin{equation}
            \xi(z) = \left[ \Omega_{k}(z) + \Omega_{R}(z) + \Omega_{m}(z) + \Omega_{\textrm{DE}}(z) \right]^{1/2} \, .
        \end{equation}

\noindent
The energy-density components for spatial curvature, radiation, and matter evolve with redshift according to their present-day values: $\Omega_k(z) = (1 + z)^2 \Omega_{k,0}$, $\Omega_m(z) = (1 + z)^3 \Omega_{m,0}$, and $\Omega_R(z) = (1 + z)^4 \Omega_{R,0}$. To implement a time-evolving DE model in \texttt{Cholla}, we adopt the Chevallier–Polarski–Linder (CPL) parameterization \citep{CPL-1, CPL-2}, in which the DE density evolves as        

        \begin{equation}
            \Omega_{\textrm{DE}}(z) = (1+z)^{3(1 + w_0 + w_a)} \exp \left[\frac{-3 w_a z}{1+z} \right] \Omega_{\textrm{DE},0} \,,
        \end{equation}
        
        \noindent The DE equation of state is described by two parameters, $w_0$ and $w_a$, defining a linear evolution with scale factor: $w_{\textrm{DE}}(a) = w_0 + w_a (1 - a)$. The case $(w_0, w_a) = (-1, 0)$ recovers the $\Lambda$CDM cosmological constant scenario, in which the DE density remains constant with expansion, $\Omega_{\textrm{DE}} = \Omega_\Lambda$. Alternative parameterizations have been proposed in the literature \citep{2025PhRvD.111b3512C, 2025IJMPD..3450061P, 2025arXiv250322529N, 2025PDU....4801912R}. We adopt the CPL formalism because it offers an intuitive first-order extension to a non-cosmological-constant DE and allows for direct comparison with the DESI Collaboration results.
        
    \subsection{Review of DDE Cosmologies} \label{subsec:sims-var}
    
	For context in interpreting our simulation results, in this section we review some key properties of the spatially flat DDE cosmologies inferred by the DESI Collaboration from their DR1 data \citep{DESI-DR1-Cosmo}. The DR1 cosmological parameters lie within 1$\sigma$ of those reported in DR2. To constrain DDE models, the DESI Collaboration combines cosmological data with three supernova samples: Pantheon+ (hereafter Pantheon) \citep{2022ApJ...938..110B}, Union3 \citep{2025ApJ...986..231R}, and DESY5 \citep{2024ApJ...973L..14D}. The resulting equations of state for all three DDE models converge at $z \approx 0.5$, where the DESI+CMB+$\Lambda$CDM cosmology is more tightly constrained (see Equation 5.9 of \cite{DESI-DR1-Cosmo}).

The cosmological parameters adopted in our simulations are listed in Table~\ref{tab:table1}. The DESI+CMB+$\Lambda$CDM cosmology provides a $\Lambda$CDM reference model derived solely from BAO and CMB measurements, without including supernova datasets. To isolate the impact of DDE, we run matched comparison simulations with a cosmological constant equation of state, $(w_0, w_a) = (-1, 0)$, using the same values of $\Omega_{m,0}$, $\Omega_{b,0}$, and $H_0$ as inferred from each SN dataset. In Table~\ref{tab:table1}, these $\Lambda$CDM comparison runs are denoted with a $\Lambda$ prefix preceding the SN dataset name.
To highlight the physical consequences of a dynamical dark energy component, we compare each DDE cosmology against its corresponding cosmological constant case. For consistency in the growth of structure, we fix the present-day matter energy density and baryon energy density in both models. This is equivalent to matching the amplitude of density fluctuations on $8\,h^{-1}\,\mathrm{Mpc}$ scales, characterized by $\sigma_8$. With these choices, any differences in baryonic structure between the DDE and $\Lambda$CDM runs reflect the effects of time-varying dark energy.

        \begin{figure*}[t]
            \centering
            \includegraphics[scale=0.402]{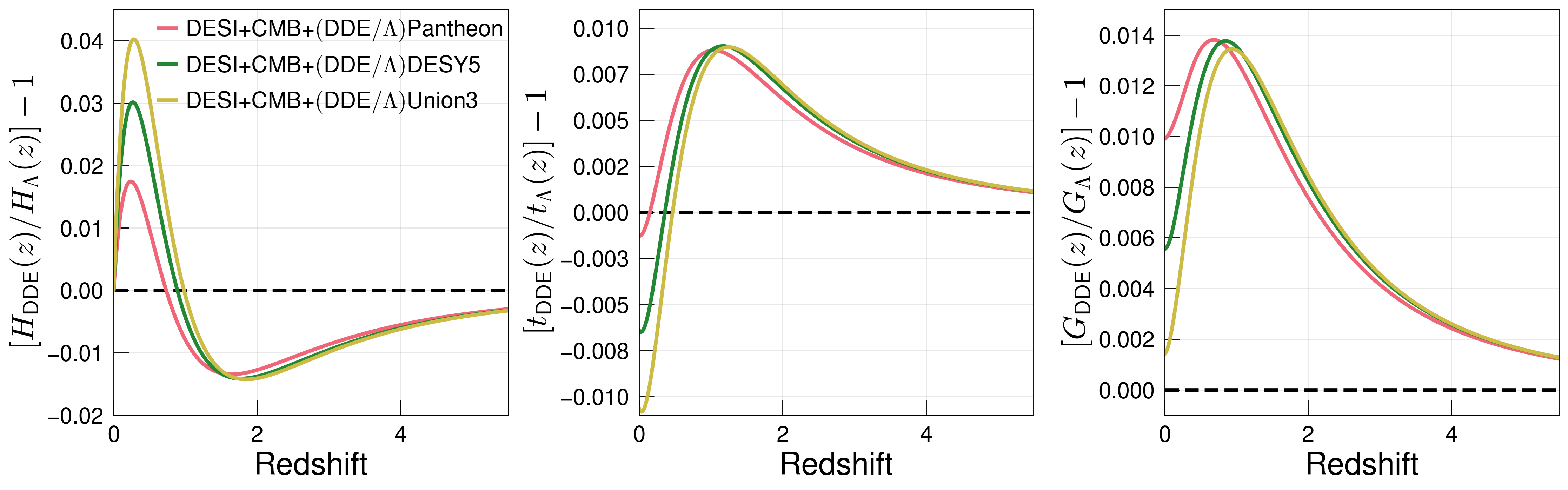}
            \caption{Relative expansion rate, age, and growth factor for a $w_0 w_a$CDM cosmology compared to their $\Lambda$CDM counterparts. 
            \emph{Left}: Taking the Hubble rate as a function of redshift, we calculate $[H_{\textrm{DDE}}(z)/H_{\Lambda}(z)] - 1$. The peach line shows this function for the Pantheon cosmologies, the green line for DESY5, and the yellow for Union3. Since the same present-day Hubble parameter $H_0$ is adopted in all comparisons, $H_{\textrm{DDE}}(0)/H_{\Lambda}(0)=1$ by construction.
            \emph{Center}: Cosmic age as a function of redshift for the DDE models against their $\Lambda$CDM counterparts. Same color coding as in the left panel. All three DDE models result in a Universe that is younger at the present epoch compared to their $\Lambda$CDM counterparts ($[t_{\textrm{DDE}}(z=0) / t_{\Lambda}(0)] < 1$). 
            \emph{Right}: Relative linear growth factor $[G_{\textrm{DDE}}(z)/G_{\Lambda}(z)] - 1$ compared to their $\Lambda$CDM counterpart models. Same color coding as in the left panel. The growth factors are not normalized at a common redshift. Since the same $\Omega_{m,0}$ is used when comparing two cosmologies, differences in growth arise solely from the DDE component. }  
	\label{fig:cosmo-info}
	\end{figure*}

        A simulation slice for the fiducial $\Lambda$CDM cosmology (denoted DESI+CMB+$\Lambda$CDM in Table~\ref{tab:table1}) is shown in Figure~\ref{fig:slab-viz}. The hierarchical nature of structure formation is evident: dark matter defines the backbone of the cosmic web, and the gas density traces the distribution of dark matter particles. The right panel shows that the highest temperatures occur in dense regions, particularly at the nodes and along the filaments of the web.
        
Differences in cosmological parameters between $\Lambda$CDM and DDE models are expected to imprint themselves on baryonic features during structure formation. To build intuition for the physical implications of these differences, we compute cosmological diagnostics for each model and present the results in Figure~\ref{fig:cosmo-info}. The left panel of Figure~\ref{fig:cosmo-info} shows the relative difference in the Hubble rate, $H_{\rm DDE}(z)/H_{\Lambda}(z) - 1$, as a function of redshift. All three DDE cosmologies exhibit a consistent trend, with qualitatively different behaviors above and below $z = 1$. At redshifts $z > 1$, the expansion rate in each DDE model is lower than in the corresponding $\Lambda$CDM case, reaching a local minimum of approximately $-1.5\%$ at $z \approx 1.5$.

       The lower expansion rate at earlier times implies that, in DDE models, diffuse gas undergoes slower adiabatic cooling, which tends to increase the Jeans mass. The opposite behavior occurs at late times. In the recent Universe ($z < 1$), the relative Hubble rate increases rapidly, reaching a maximum deviation of 1.5\%, 3\%, and 4\% at $z \approx 0.25$ for the DESI+CMB+Pantheon, DESI+CMB+DESY5, and DESI+CMB+Union3 DDE cosmologies, respectively, compared to their $\Lambda$CDM counterparts. At lower redshifts ($z < 0.25$), the relative difference in the Hubble rate declines rapidly, converging to the same present-day value $H_0$ in both DDE and $\Lambda$CDM models.
        
        The different scale factor evolution rates in DDE and $\Lambda$CDM cosmologies correspond to different age–redshift relations. To illustrate how the cosmic expansion history diverges with redshift, we plot the relative age differences given by Equation~\ref{eqn:age} in Figure~\ref{fig:cosmo-info}. All three DDE models exhibit a similar trend in relative age difference compared to their $\Lambda$CDM counterparts. At high redshift ($z > 5$), DDE cosmologies are approximately 0.1\% older. This slightly older age implies marginally more time for adiabatic cooling of the diffuse gas. The relative age difference increases with decreasing redshift, reaching a peak of 0.8\% at $z \approx 1$, coincident with the redshift at which the Hubble rate in DDE models begins to exceed that in $\Lambda$CDM. Below $z \approx 1$, the relative age difference declines steeply, reaching present-day values of $-0.1$\%, $-0.6$\%, and $-1$\% for the DESI+CMB+Pantheon, DESI+CMB+DESY5, and DESI+CMB+Union3 DDE models, respectively. Given an identical redshift-dependent UV background, a younger DDE Universe would undergo less integrated photoheating, which may influence the temperature distribution of the IGM at late times.

        The distribution of large-scale structure can be modeled using linear perturbation theory, while non-linear processes dominate on smaller scales. In the $\Lambda$CDM framework, the evolution of linear density fluctuations is primarily governed by the matter density parameter, $\Omega_{m,0}$. In DDE cosmologies, the time-dependent dark energy equation of state alters the evolution of the dark energy–to–matter energy-density ratio, which can in turn modulate the rate of structure growth. To quantify this effect, we compute the linear growth factor $D_{+}(z)$ by numerically integrating the linear density perturbation equation. We normalize the result by the scale factor to define the dimensionless growth function, $G(z) \equiv D_{+}(z)/a(z)$, following the formalism of \citet{DDE-ODE-Linder}

        \begin{equation} \label{eqn:ODE-DDE}
		G'' + \left[\frac{7}{2} - \frac{3}{2} \frac{w_{\textrm{DE}}(a)}{1 + X(a)} \right] \frac{G'}{a} + \frac{3}{2} \frac{1 - w_{\textrm{DE}}(a)}{1 + X(a)} \frac{G}{a^2} = 0 \,,
	\end{equation}
	
	\noindent where $X(a) = \Omega_{m}(a) / \Omega_{\textrm{DE}}(a)$ is the ratio of matter to dark energy density as a function of the scale factor. The evolution of the dimensionless linear growth function $G(a) = D_{+}(a)/a$ is shown as a function of redshift in the right-most panel of Figure~\ref{fig:cosmo-info}.

	Since the influence of dark energy is strongest at low redshift, the linear growth factor is expected to converge across models in the early Universe. We find that all DDE cosmologies considered exhibit a slightly enhanced linear growth factor relative to their $\Lambda$CDM counterparts at all redshifts. At $z \approx 5$, the DDE models show a $\sim$0.2\% higher growth factor compared to the corresponding model with $(w_0, w_a) = (-1, 0)$. This relative difference increases toward lower redshift, reaching a maximum of approximately 1.4\% at $z = 0.68$, $0.84$, and $0.94$ for the DESI+CMB+Pantheon, DESI+CMB+DESY5, and DESI+CMB+Union3 models, respectively. By $z = 0$, the linear growth factors in these DDE cosmologies exceed their $\Lambda$CDM analogs by 1.0\%, 0.5\%, and 0.1\%. All three models follow a qualitatively similar trajectory, indicating that DDE scenarios may lead to a modest enhancement in the amplitude of large-scale structure.

	\citet{DDE-PowerSpectra} compare matter power spectra by matching the comoving distance to the surface of last scattering in a $w_0w_a$CDM cosmology to that of a model with an effective constant equation of state ($\mathrm{d}w/\mathrm{d}a = 0$). For the specific case of $(w_0, w_a) = (-0.8, -0.732)$, they find that the linear growth factor is lower than in the $w = -1$ model up to $z = 3.12$, with the largest relative suppression occurring near $z \approx 1$. Moreover, they show that DDE models with negative $w_a$ values predict enhanced small-scale power ($k \gtrsim 1\ h,\mathrm{Mpc}^{-1}$), with deviations of up to 2\% in the matter power spectrum at $k = 3\ h,\mathrm{Mpc}^{-1}$ after removing the effect of linear growth. 
    
    These results suggest that when the linear growth factor is enhanced in a DDE cosmology, $[G_{\textrm{DDE}}(z)/G_{\Lambda}(z)] > 1$, the resulting increase in structure formation may lead to stronger shock and compressional heating of the IGM. While distinct from photoheating by the UV background, this additional gravitational heating could modify the thermal evolution of the diffuse gas.

Since a DDE component primarily alters the expansion history of the Universe, any growth-related observable that depends on the Hubble rate may be affected. As shown in Figure~\ref{fig:cosmo-info}, all DDE models exhibit a lower Hubble rate relative to $\Lambda$CDM at $z \gtrsim 1$. This reduced expansion rate leads to slightly older cosmic ages at fixed redshift, allowing more time for adiabatic cooling of the intergalactic gas and greater cumulative energy injection from the photoionizing UV background. In parallel, differences in the linear growth factor influence the efficiency of gravitational heating processes such as shocks and compression. The interplay between adiabatic cooling, photoheating, and structure-driven heating can alter the thermal state of the IGM. In this study, we investigate how these competing processes shape the temperature and ionization structure of the IGM as probed by the \Lya\ forest.

    \begin{figure*}
            \centering
            \includegraphics[scale=0.3925]{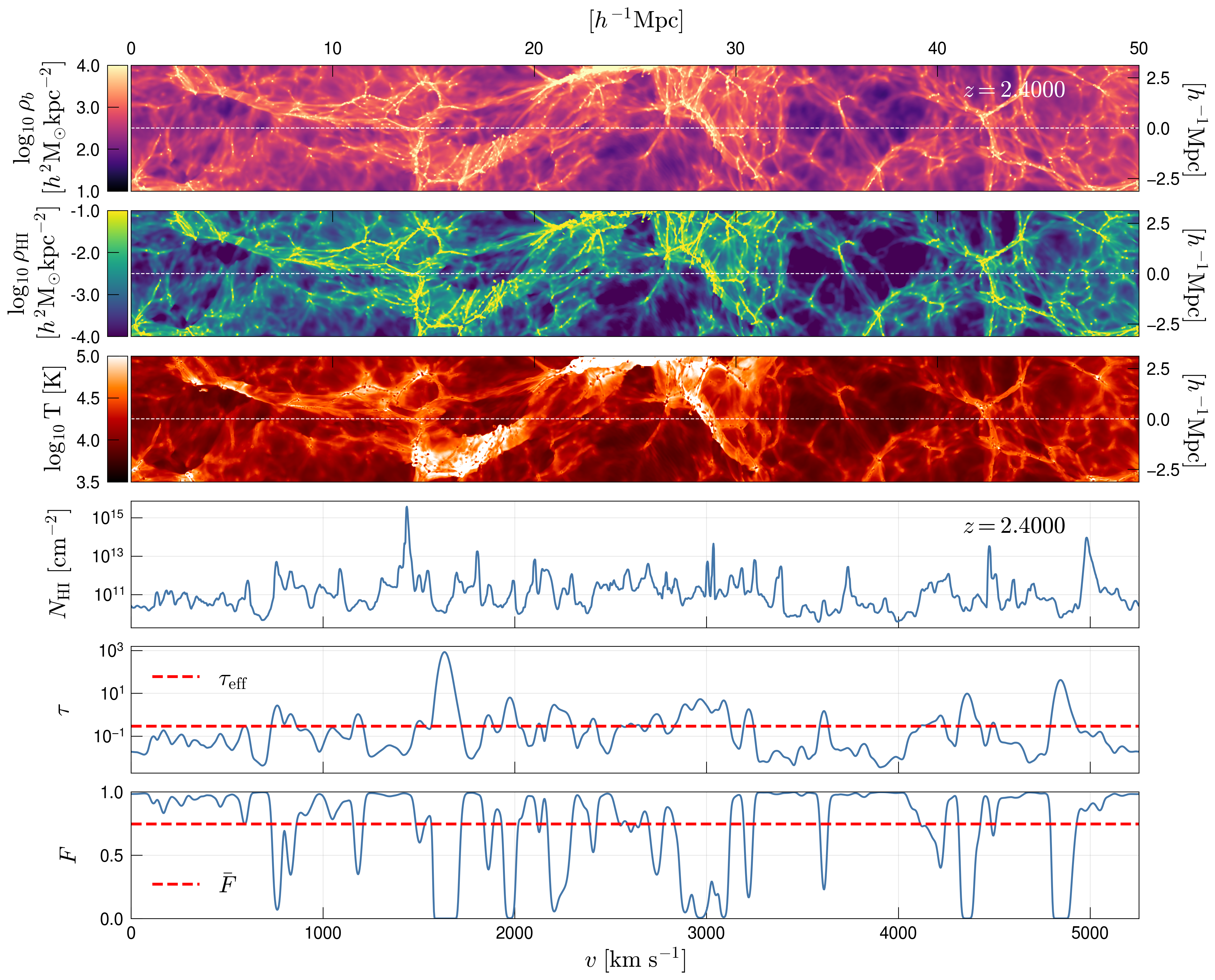}
            \caption{Example of a synthetic \Lya\ spectrum at redshift $z=2.4$ for the fiducial DESI+CMB+$\Lambda$CDM cosmology. The top two panels display the surrounding gas and \HI\ density, projected along $10^{3}\ h^{-1}\textrm{kpc}$. The third panel displays a temperature projection weighted by the gas density. The fourth panel shows physical column density $N_{\textrm{HI}}$. The bottom two panels display the optical depth and transmitted flux along the skewer. The mean flux along the LOS $\bar{F}$ defines the effective optical depth $\tau_{\textrm{eff}} = -\log\left(\bar{F}\right)$ for this skewer.}
            \label{fig:skew-viz}
        \end{figure*}

    \subsection{Optical Depth}\label{subsec:opt-depth}

        To study the \Lya\ forest, we generate approximately $7.86 \times 10^{5}$ synthetic spectra per redshift, corresponding to $3 \times (2048 / 4)^2$  LOS extracted along the three principal axes of the simulation volume. For each LOS, we compute the neutral hydrogen optical depth and the corresponding transmitted flux in all simulation cells. The optical depth is calculated as the product of the hydrogen photoabsorption cross section and the local number density of \HI\ along the LOS,

        \begin{equation}
            \tau_\nu = \int n_{\textrm{HI}} \sigma_\nu \textrm{d}r \,.
        \end{equation}

        \noindent This calculation is performed in redshift space, accounting for peculiar velocities, so that $\mathrm{d}r$ corresponds to the physical path length along the line of sight. We approximate the absorption line shape using a Doppler profile, retaining only the Gaussian core of the full Voigt profile. Under this approximation, the \Lya\ scattering cross section takes the form
        
        \begin{equation}
            \sigma_\nu = \frac{\pi e^2}{m_e c} f_{12} \frac{1}{\Delta \nu_D}  \frac{1}{\pi^{1/2}} \exp\left(-x^2 \right) \,,
        \end{equation}

        \noindent where $e$ is the elementary charge, $m_e$ is the electron mass, and $f_{12}$ is the \Lya\ oscillator strength. The Doppler broadening $\Delta \nu_D$ arises from the thermal motion of the gas and is given by $\Delta \nu_D = (v_{\rm th} / c), \nu_0$, where $v_{\rm th} = \sqrt{2k_{\rm B}T/m_{\rm H}}$ is the thermal velocity of hydrogen and $\nu_0$ is the Ly$\alpha$ line-center frequency. The exponential term depends on the dimensionless frequency shift $x = (\nu - \nu_0)/\Delta \nu_D$. A photon emitted at frequency $\nu_0$ from a cell with peculiar velocity $u_0$ will be resonantly scattered by gas in a cell moving with velocity $u$ when its Doppler-shifted frequency matches the local \Lya\ line center:

        \begin{equation}
            \nu = \nu_0 \left(1 + \frac{u_0 - u}{c} \right) \,.
        \end{equation}

        \noindent The \Lya\ cross section as a function of physical velocity is then

        \begin{equation}
            \sigma_{u_0} = \frac{\pi e^2}{m_e c} f_{12} \frac{\lambda_0}{v_{\textrm{th}}}  \frac{1}{\pi^{1/2}} \exp\left[-\left(\frac{u_0 - u}{v_{\textrm{th}}} \right)^2 \right] \,,
        \end{equation}
        
        \noindent where $\lambda_0 = 1216\ \textrm{\AA}$ is the rest-frame wavelength of the \Lya\ feature.

        When integrating along the LOS of a skewer to compute the optical depth at a given velocity $u_0$, we convert comoving coordinates to physical units using $\mathrm{d}r = a,\mathrm{d}x$. Applying the definition of the Hubble parameter, $H = \dot{a}/a$, and recognizing that in redshift space $\mathrm{d}r = \mathrm{d}u/H$ (with $u$ the peculiar velocity coordinate), we arrive at the expression:

        \begin{equation}\label{eq:tau_u0}
            \tau_{u_0} = \frac{\pi e^2}{m_e c} f_{12} \frac{\lambda_0}{H} \int  \frac{n_{\textrm{HI}}}{v_{\textrm{th}} \pi^{1/2}}   \exp\left[-\left(\frac{u - u_0}{v_{\textrm{th}}} \right)^2 \right] \textrm{d}u \,.
        \end{equation}

        \noindent We measure the optical depth along a skewer by setting $u_0$ as the cell-centered Hubble flow velocity. The integral contribution from the $i$-th cell to the optical depth at the $j$-th cell becomes

        \begin{equation}
             n_{\textrm{HI},i} \int_{u_{i - 1/2}}^{u_{i + 1/2}} \frac{1}{v_{\textrm{th}, i} \pi^{1/2}}   \exp\left[-\left(\frac{u_i - u_j}{v_{\textrm{th}, i}} \right)^2 \right] \textrm{d}u_i \,,
        \end{equation}
        	
        \noindent where $n_{\textrm{HI},i}$ is the physical number density of \HI\ in the $i$-th cell, $v_{\textrm{th}, i}$ is the Doppler broadening of the gas in the $i$-th cell, $u_i$ is the cell-centered velocity we integrate over, and $u_j$ is the cell-centered Hubble flow through the $j$-th cell.

        Analytically solving the integral (see Appendix \ref{app:opt-depth}), we find that the optical depth for the $j$-th cell along the LOS is
                
        \begin{equation}\label{eq:opt-depth-discrete}
            \tau_j = \frac{\pi e^2 f_{12} \lambda_0}{m_e c H} \sum_i  \frac{n_{\textrm{HI},i}}{2} \left[\textrm{erf}(y_{\textrm{R},i}) - \textrm{erf}(y_{\textrm{L},i})\right] \,,
        \end{equation}

        \noindent where we sum the contribution from all other cells $i$ along the skewer. This method of analysis follows from previous studies including \cite{2021ApJ...912..138V} and \cite{2015MNRAS.446.3697L}. The error function argument is the difference between the cell interface Hubble flow and the gas velocity. For the right interface, the argument is $y_{\textrm{R},i} = [v_{\textrm{R},H,j} - (v_{\textrm{C},H,i} + u_i)] / v_{\textrm{th}, i}$, where $v_{\textrm{R},H,j}$ is the Hubble flow along the right interface of the $j$-th cell, $v_{C,H,i}$ is the cell centered Hubble flow of the $i$-th cell, $u_i$ is the peculiar velocity of the gas in the $i$-th cell, and $v_{\textrm{th},i}$ is the Doppler broadening of the gas in the $i$-th cell. Likewise, $y_{\textrm{L},i} = [v_{\textrm{L},H,j} - (v_{\textrm{C},H,i} + u_i)] / v_{\textrm{th}, i}$ takes into account the Hubble flow along the left interface of the $j$-th cell. 
        
      A representative synthetic \Lya\ spectrum is shown in Figure~\ref{fig:skew-viz}. The neutral hydrogen column density along the line of sight spans a broad range, from $10^{10}$ cm$^{-2}$ to $10^{15}$ cm$^{-2}$. Regions of high column density correspond to strong \Lya\ absorption features, characterized by large optical depths and minimal transmitted flux. The skewer intersects the densest structure along the LOS at a velocity coordinate of approximately 1500 km s$^{-1}$. Due to the peculiar velocity field within the filament, the peak in $N_{\textrm{HI}}$ is shifted in redshift space, producing maximum absorption (i.e., zero transmitted flux) near 1800 km s$^{-1}$. In the velocity interval 3100--4000 km s$^{-1}$, the skewer passes through a large underdense region, interrupted by a filament that produces a prominent absorption feature around 3600 km s$^{-1}$. In the surrounding void, the high transmitted flux reflects the low neutral hydrogen density and the consequent transparency of the IGM to \Lya\ photons.

\section{Results}\label{sec:res}

    The three DDE cosmological models examined in this work exhibit distinct Hubble rate and cosmic age evolutions compared to their $\Lambda$CDM counterparts. Since DE primarily affects the large-scale expansion of the Universe, its influence is expected to manifest most clearly in low-density environments. At the same time, differences in the linear growth factor may alter the formation and evolution of nonlinear structures, potentially modifying the physical conditions in overdense regions. The results of our simulations are presented as follows. In Section~\ref{subsec:IGM-temp}, we examine the density and temperature distribution of the IGM. Section~\ref{subsec:Lya-tau} presents the evolution of the hydrogen optical depth. In Section~\ref{subsec:Lya-FPS}, we compute the \Lya\ forest flux power spectrum (FPS), and in Section~\ref{subsec:Lya-FPS-tilt}, we highlight our primary result: a spectral tilt in the FPS induced by the DDE models. Finally, Section~\ref{subsec:Lya-FPS-pec} explores the effect of peculiar velocities on the FPS by comparing calculations in real and redshift space.
    
    \subsection{IGM Density and Temperature}\label{subsec:IGM-temp}

\begin{figure}
            \centering
            \includegraphics[scale=0.455]{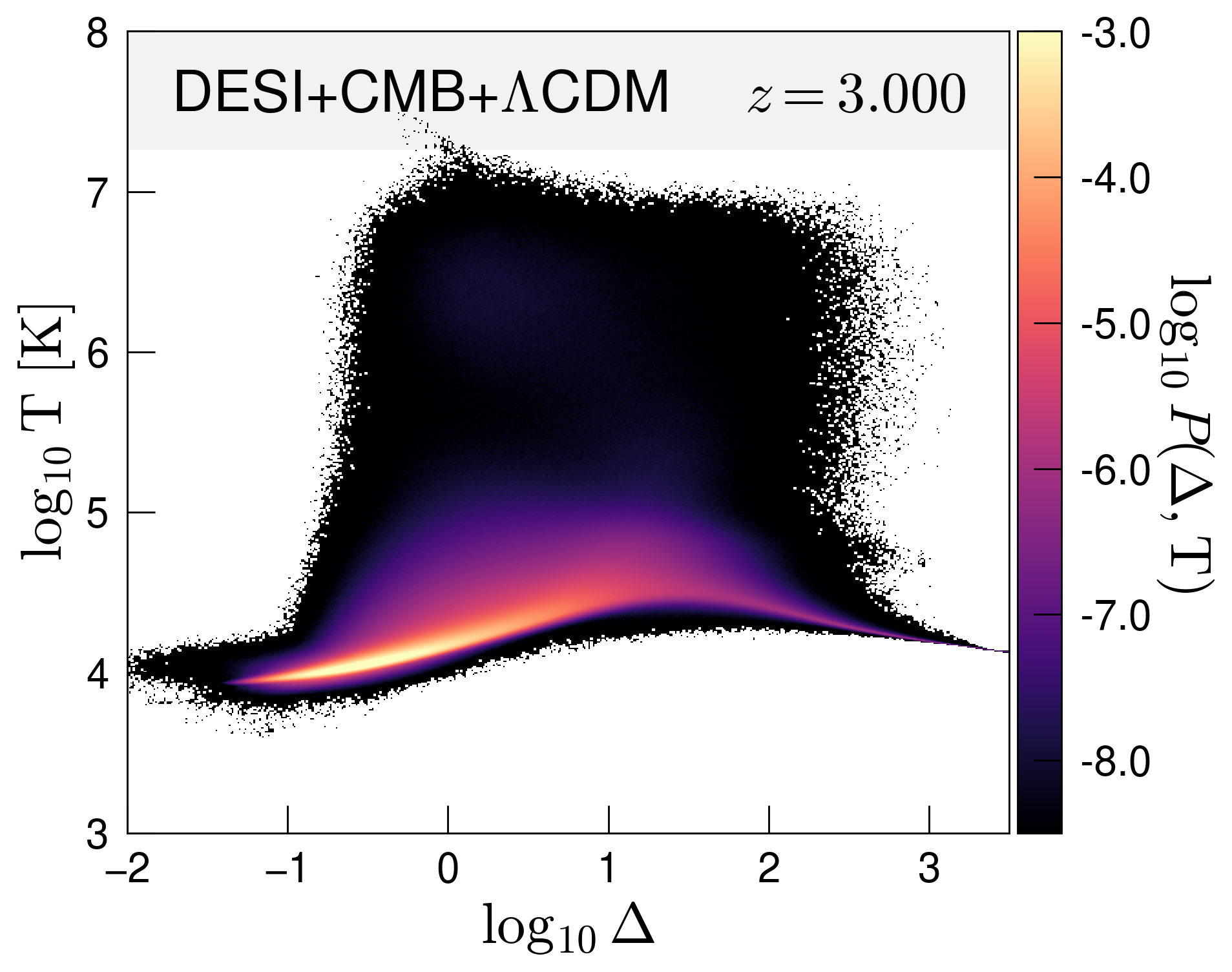}
            \caption{Distribution of gas overdensity $\rho / \bar{\rho}_b$ and temperature at $z = 3.0$ for the fiducial $\Lambda$CDM cosmology. The color scale indicates the logarithmic probability that a simulation cell falls within a given bin of the two-dimensional $\Delta$--$T$ volume-weighted histogram. The interstellar medium cool phase, where the colder/denser gas within galaxies are at $\Delta>2$, is not resolved in our simulations. White regions correspond to bins with no occupied cells, i.e., $P(\Delta, T) = 0$.}
            \label{fig:density-temp}
        \end{figure}

	The thermal state of the IGM is governed by a complex interplay between various cooling and heating processes. Cooling mechanisms include free-free Bremsstrahlung, collisional excitation, collisional ionization, inverse Compton scattering, and recombination, while heating is primarily driven by the UV background associated with cosmic reionization. Adiabatic expansion of the Universe introduces an additional cooling channel and influences the large-scale gas density distribution. To examine the temperature and density of the gas in our simulations, we present the $\Delta$--$T$ phase diagram in Figure~\ref{fig:density-temp} for the fiducial $\Lambda$CDM cosmology. Here, $\Delta$ denotes the overdensity relative to the mean cosmic baryon density, $\bar{\rho}_b$. Most gas cells fall within the density range $-1 < \log_{10} \Delta < 2$. A non-negligible fraction occupies the temperature range characteristic of the diffuse warm-hot intergalactic medium (WHIM), with $5 < \log_{10} (T/{\rm K}) < 7$. The highest overdensities, $\log_{10} \Delta > 2$, correspond to the densest regions, and sites of potential galaxy and galaxy cluster formation.
        
    \begin{figure*}[t]
    \centering
    \includegraphics[scale=0.402]{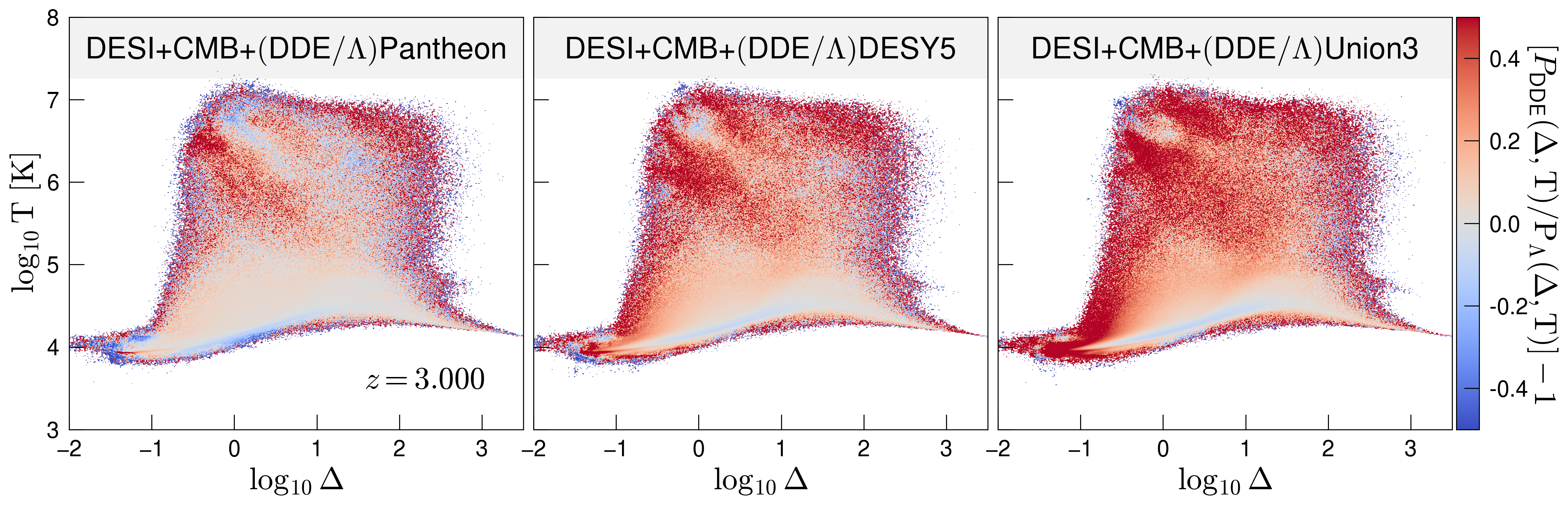}
    \caption{Relative difference in the phase-space occupancy of gas cells between $w_0w_a$CDM cosmologies and their corresponding $\Lambda$CDM counterparts at redshift $z = 3$. The quantity plotted is the fractional difference in the joint probability distribution: $[P_{\textrm{DDE}}(\Delta, T) / P_{\Lambda}(\Delta, T)] - 1$, computed in $\log_{10} \Delta$--$\log_{10} T$ space. From left to right, the panels show the results for the DESI+CMB+Pantheon, DESI+CMB+DESY5, and DESI+CMB+Union3 models. Red (blue) regions indicate where a phase-space bin is more (less) populated in the DDE model than in its $\Lambda$CDM counterpart.}
    \label{fig:diff-density-temp}
\end{figure*}
        
    We present the relative differences in the $\Delta$--$T$ phase space for the $w_0w_a$CDM cosmologies in Figure~\ref{fig:diff-density-temp}. The predominantly red coloring at temperatures $T > 10^5$ K indicates that each DDE model produces a systematically warmer IGM compared to its corresponding $\Lambda$CDM counterpart. In high-density regions with $\log_{10} \Delta > 2$, the absence of a dominant color gradient suggests that the density–temperature distribution remains similar across cosmologies. Around $(\log_{10} \Delta, \log_{10} T) = (0, 6)$, we observe a clear overdensity of cells in all $w_0w_a$CDM models relative to $\Lambda$CDM, reflected by an excess in $[P_{\textrm{DDE}}(\Delta, T)/P_{\Lambda}(\Delta, T)] - 1$. This excess is largest in the model with the greatest deviation from $\Lambda$CDM in its expansion history, despite the low overall occupancy in this region ($\log_{10} P(\Delta, T) < -6$). In contrast, distinct diagonal blue features indicate that cells near mean density ($\Delta \approx 1$) are more frequently populated in the $\Lambda$CDM models. Overall, DDE cosmologies preferentially populate warmer and more diffuse regions in the $\Delta$--$T$ phase space.

     The density–temperature relationship of gas in the IGM is traditionally approximated by a power-law relation of the form $T(\Delta) = T_0\, \Delta^{\gamma - 1}$, where $T_0$ is the characteristic temperature at mean density and $\gamma$ describes the slope of the  relation around the cosmic mean \citep{1997MNRAS.292...27H, 2016ARA&A..54..313M, 2015MNRAS.450.4081P}. The effect of a UV ionizing background under non-equilibrium ionization conditions on the power-law temperature–density relation was studied by \cite{Puchwein-15}, who emphasized the ambiguity in defining a unique IGM temperature. More recently, high-resolution hydrodynamic cosmological simulations \cite{2022ApJ...933...59V} show that the most probable temperature can deviate from $T_0$ by up to 15\% in the vicinity of the cosmic mean density, specifically within the range $-1 \leq \log_{10} \Delta \leq 1$.

      Rather than characterizing the IGM temperature across a broad dynamic range in overdensity, we focus our analysis on simulation cells within half an order of magnitude of the cosmic mean baryon density and exclude the hottest gas phases. To trace the thermal evolution of the diffuse IGM in this regime, we compute the expectation value of the temperature from the normalized probability distribution

\begin{equation}
P(T) = \int_{-0.25}^{0.25} P(\Delta, T|3 < \log_{10}(T/{\rm K}) < 5) \mathrm{d}\log_{10} \Delta,
\end{equation}

\noindent and define the resulting characteristic temperature as the low-density IGM temperature, $T_{\textrm{LD}}$. In other words, $T_{\textrm{LD}}$ is the weighted average of the gas temperature within the range $-0.25 \leq \log_{10} \Delta \leq 0.25$, restricted to gas cooler than the WHIM, i.e., $10^3\,\mathrm{K} \leq T \leq 10^5\,\mathrm{K}$.

    \begin{figure}
    \centering
    \includegraphics[scale=0.456]{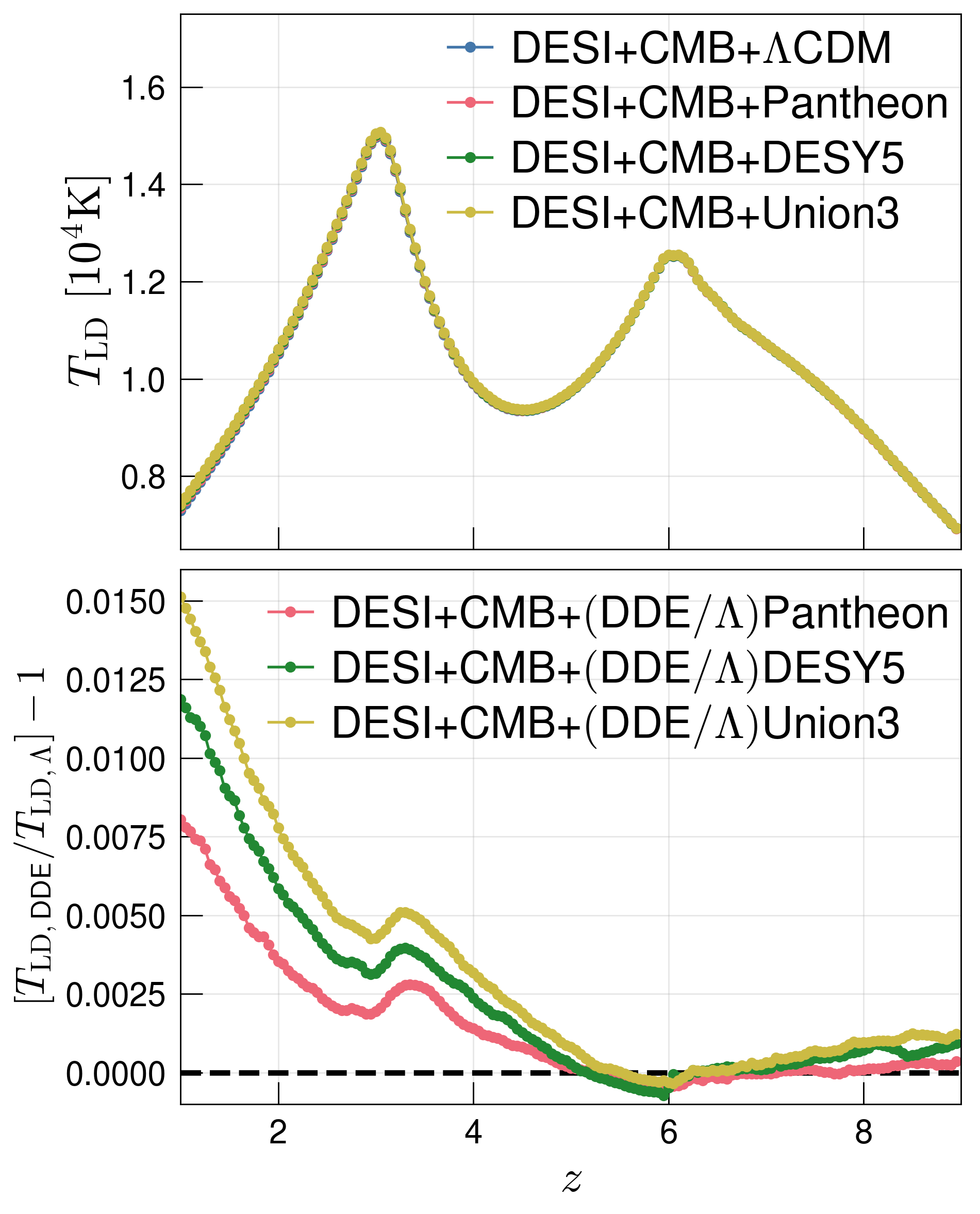}
    \caption{Thermal history of the IGM in all three $w_0w_a$CDM cosmologies and the fiducial $\Lambda$CDM model, along with the relative temperature difference between each DDE model and its corresponding $\Lambda$CDM counterpart.
    \emph{Top}: Evolution of the low-density IGM temperature, $T_{\textrm{LD}}$, for all cosmologies considered, with parameters inferred from the DESI Collaboration. The yellow curve corresponds to the DESI+CMB+Union3 cosmology; the blue curve represents the fiducial $\Lambda$CDM baseline; peach and green curves correspond to the DESI+CMB+Pantheon and DESI+CMB+DESY5 cosmologies, respectively. The peaks at $z \approx 6$ and $z \approx 3$ reflect the thermal impact of hydrogen and helium reionization. The temperature histories across models are nearly indistinguishable, with only minor deviations.
    \emph{Bottom}: Relative difference in $T_{\textrm{LD}}$ between each $w_0w_a$CDM cosmology and its $\Lambda$CDM counterpart. Values of $[T_{\textrm{LD,DDE}} / T_{\textrm{LD},\Lambda}] > 1$ indicate that the low-density IGM is warmer in the DDE cosmology.}
    \label{fig:temp-history}
\end{figure}

        The IGM temperature histories for each DDE cosmology are shown in Figure~\ref{fig:temp-history}. A period of monotonic temperature increase characterizes the early Universe up to the completion of hydrogen reionization around $z \approx 6$. Following this, adiabatic expansion cools the diffuse, low-density IGM until a temperature minimum is reached near $z \approx 4.5$, after which a second heating phase begins due to extreme UV photons from active galactic nuclei (AGN). The subsequent temperature rise until $z \approx 3$ corresponds to the epoch of helium reionization, during which singly ionized helium (HeII) is fully ionized. After this period, continued adiabatic expansion drives a second phase of cooling. All $w_0w_a$CDM and $\Lambda$CDM cosmologies exhibit similar thermal evolution, broadly following the same trends.

As illustrated in the top panel of Figure~\ref{fig:temp-history}, the thermal history of the IGM is primarily governed by the competition between adiabatic cooling from cosmic expansion and photoheating from the UV background. Since all models are exposed to the same redshift-dependent metagalactic UV flux, any differences in thermal history arise from variations in the expansion rate. A deviation from the $\Lambda$CDM expansion history at fixed redshift modifies the adiabatic cooling rate, which in turn could imprint subtle differences in the mean neutral hydrogen density. However, because the DDE cosmologies broadly track the same expansion behavior, we expect only minimal deviations in the low-density IGM temperature, $T_{\textrm{LD}}$, across models.

        The difference in IGM temperature evolution between each $w_0w_a$CDM cosmology and its corresponding $\Lambda$CDM model is shown in the bottom panel of Figure~\ref{fig:temp-history}. Prior to the completion of hydrogen reionization at $z \approx 6$, the IGM is slightly warmer in the DDE models. As the IGM enters a cooling phase dominated by adiabatic expansion, the low-density temperature in all $w_0w_a$CDM cosmologies becomes increasingly elevated relative to their $\Lambda$CDM counterparts. A local maximum in the temperature difference occurs around $z \approx 3.3$, where the $\Lambda$CDM versions are approximately 0.25–0.50\% cooler than their DDE analogs. At the end of helium reionization near $z \approx 3$, where the IGM reaches its peak temperature, this difference exhibits a local minimum. The sensitivity of the IGM thermal state to the expansion rate becomes most pronounced after this final heating phase, with $T_{\textrm{LD}}$ in the DDE models exceeding that of $\Lambda$CDM by 0.76\%, 1.23\%, and 1.5\% at $z = 1.0$ for the Pantheon, DESY5, and Union3 cosmologies, respectively. This growing divergence at late times suggests that the expansion history leaves a measurable imprint on the thermal state of the IGM during adiabatic cooling. Overall, we find that the DDE-induced modification to the low-density IGM temperature remains below the 2\% level across all redshifts and cosmologies considered.
        
    \subsection{\Lya\ Optical Depth Evolution}\label{subsec:Lya-tau}
        
\begin{figure}
    \centering
    \includegraphics[scale=0.456]{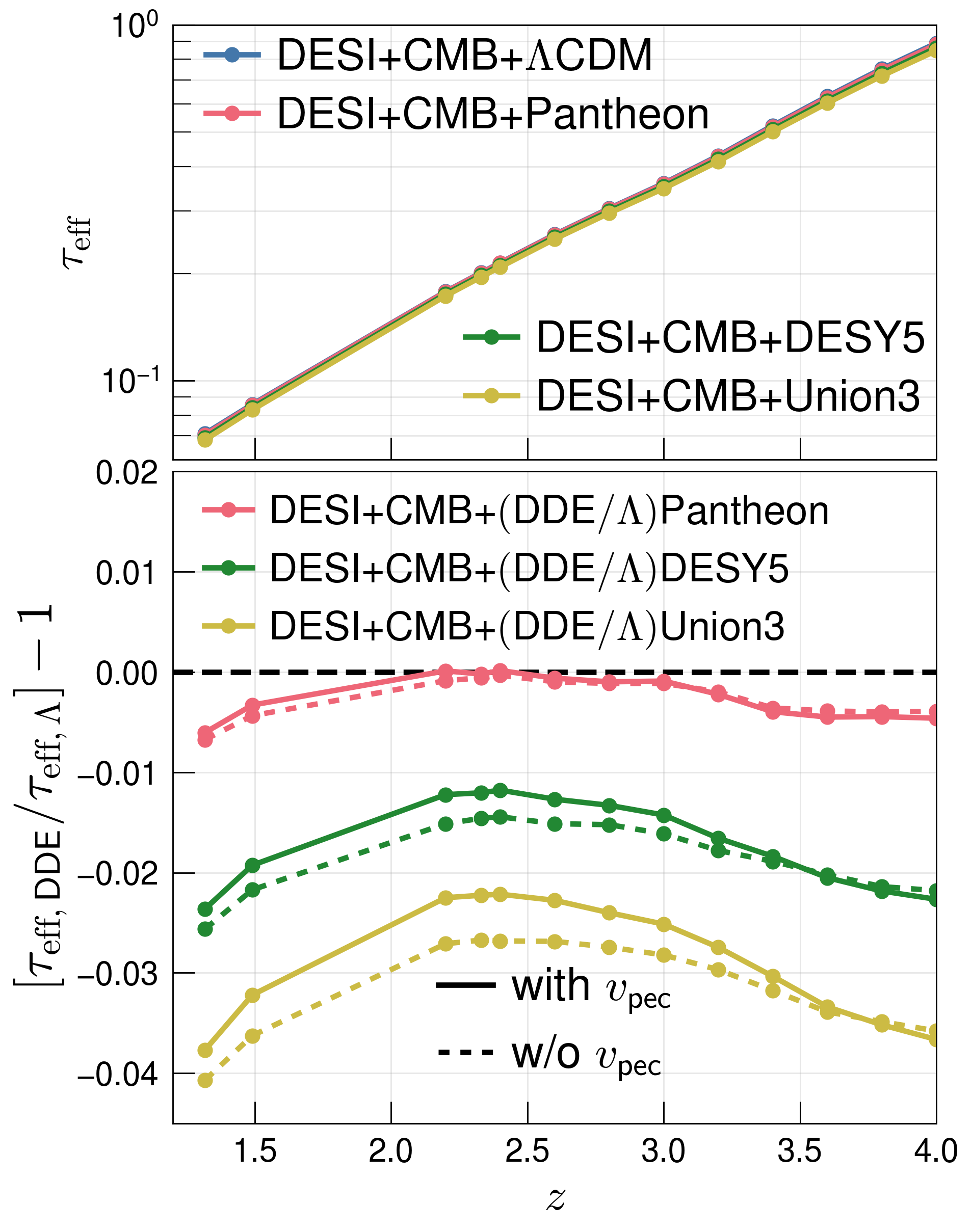}
    \caption{
        Evolution of the effective optical depth for all cosmological models considered in this study. Colors match those used in Figure~\ref{fig:temp-history}. 
        \emph{Top}: Median effective optical depth as a function of redshift. At each snapshot, we compute the distribution of $\tau_{\rm eff}$ across all skewers and plot the median value. The decline in $\tau_{\rm eff}$ reflects the continued photoionization of residual \HI\ and the progressive thinning of the IGM.
        \emph{Bottom}: Relative difference in $\tau_{\rm eff}$ for all $w_0 w_a$CDM models with respect to their $\Lambda$CDM counterparts. Dashed lines indicate the effect of excluding peculiar velocities. The IGM is systematically more optically thin in DDE cosmologies.
    }
    \label{fig:optdepth-evolution}
\end{figure}

The photoionization of \HI\ by quasar radiation imprints a forest of absorption lines on the rest-frame spectra redward of $1216\ \textrm{\AA}$. The effective optical depth of the \Lya\ forest provides a direct measure of the \HI\ content in the diffuse IGM and offers key insights into the photoionization history of the high-redshift Universe \citep{1965ApJ...142.1633G}. Using the calculations described in Section~\ref{subsec:opt-depth}, the top panel of Figure~\ref{fig:optdepth-evolution} shows the median effective optical depth from synthetic spectra, computed as the negative logarithm of the mean transmitted flux along each skewer. The median $\tau_{\rm eff}$ decreases monotonically in all simulations as the Universe expands, diluting $n_{\textrm{HI}}$ and enabling further photoionization. Similar to the temperature evolution in Figure~\ref{fig:temp-history}, all models follow the same general trend and exhibit only minor differences.
        
Because DDE models predict a warmer IGM, the neutral hydrogen fraction $x_{\textrm{HI}} = n_{\textrm{HI}} / (n_{\textrm{HI}} + n_{\textrm{HII}})$ may decline more rapidly than in $\Lambda$CDM. We compute the cell-by-cell relative difference in $x_{\textrm{HI}}$, finding that the median of the logarithmic distribution differs at the sub-percent level across all redshifts: $-0.4\% < [\log_{10}(x_{\textrm{HI, DDE}}) / \log_{10}(x_{\textrm{HI}, \Lambda})] - 1 < 0\%$. This slight suppression of the neutral fraction in DDE models remains approximately constant with redshift. Similarly, the median logarithmic difference in the neutral hydrogen density follows $-0.5\% < [\log_{10}(n_{\textrm{HI, DDE}}) / \log_{10}(n_{\textrm{HI}, \Lambda})] - 1 < 0\%$, with the largest deviations occurring at lower redshifts.

To assess the cumulative effect on \Lya\ absorption, we compute the redshift evolution of the effective optical depth $\tau_{\rm eff}$ for all models. The bottom panel of Figure~\ref{fig:optdepth-evolution} shows that $w_0w_a$CDM cosmologies predict a systematically more transparent IGM. The relative decrease in $\tau_{\rm eff}$ reaches a local minimum of $-1.2\%$ for the DESI+CMB+DESY5 cosmology and $-2.2\%$ for DESI+CMB+Union3 near $z\approx 2.25$, while the DESI+CMB+Pantheon model remains close to zero. By $z=1.317$, all DDE cosmologies converge toward the same relative difference observed at $z=4$: $-0.5\%$, $-2.2\%$, and $-3.8\%$ for Pantheon, DESY5, and Union3, respectively.

Our calculation of $\tau_{\rm eff}$ includes the effects of peculiar velocities, which introduce redshift-space distortions. Previous studies have shown that peculiar motions enhance line blending and increase the effective optical depth \citep{vpec_tau1, vepc_tau2}. We confirm that this effect is present in our models, finding $\left[\tau_{\rm eff}(v_{\rm pec}=0) - \tau_{\rm eff}\right]/\tau_{\rm eff} > 0$ across the board. The impact of peculiar velocities is largest at $z < 3.5$, where their omission most strongly alters the relative difference between DDE and $\Lambda$CDM models.

A warmer IGM leads to increased Doppler broadening of the \Lya\ line profile, smearing the absorption feature over a larger velocity width. A smoother absorption feature may redistribute the local optical depth, potentially changing the total effective optical depth. This effect becomes more apparent when redshift-space distortions are excluded, that is, when peculiar velocities along the line of sight are omitted. The declining relative difference in $\tau_{\rm eff}$ for the more optically thin $w_0 w_a$CDM cosmologies from $z\approx 2.20$ to $z\approx 1.491$ may be linked to a downturn in the hydrogen photoionization rate near $z\approx 2$, as shown in the leftmost column of Figure~9 in \cite{2022ApJ...933...59V}.
    
    	\begin{figure*}
    \centering
    \includegraphics[scale=0.457]{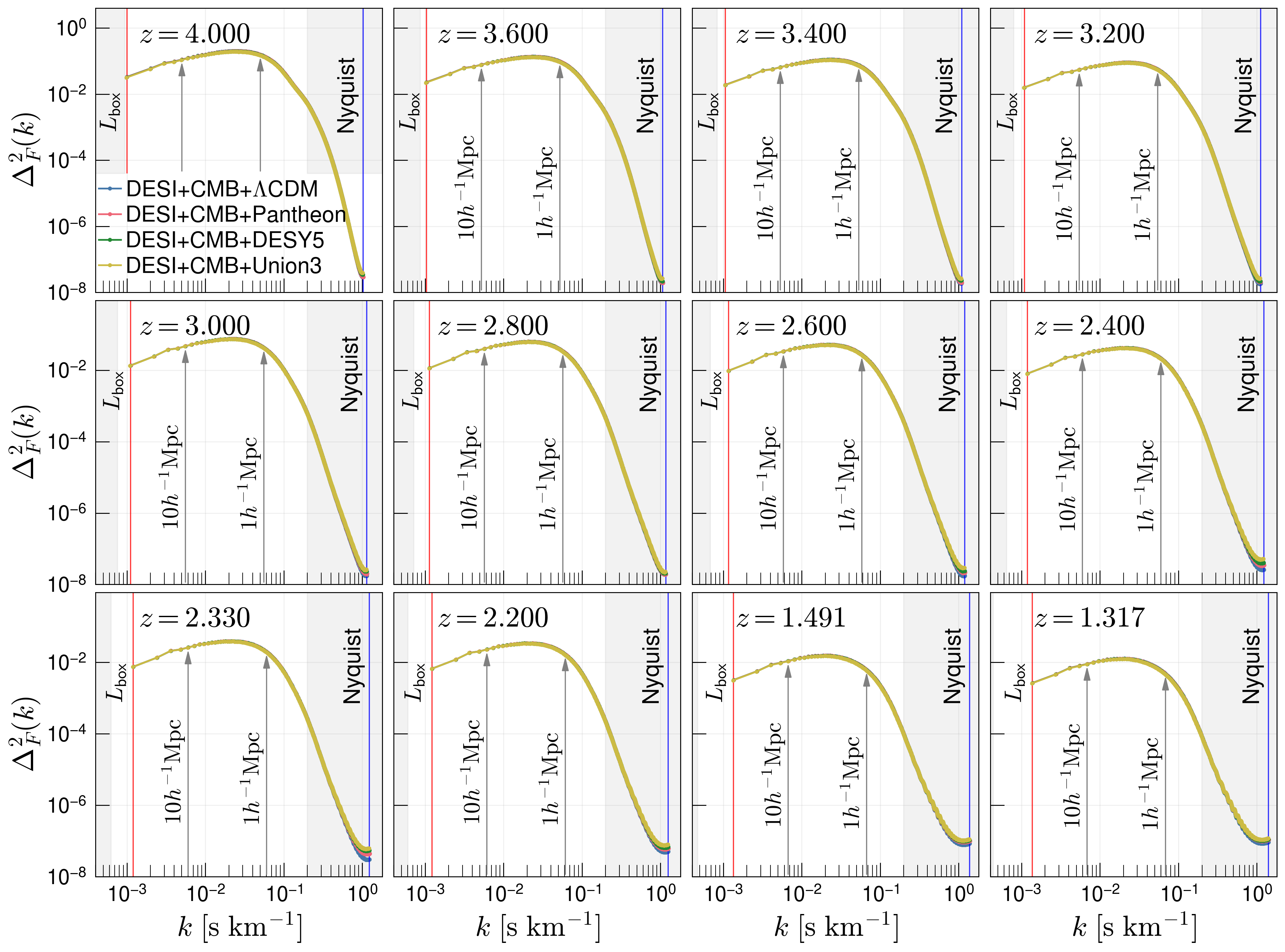}
    \caption{Evolution of the flux power spectrum across redshift for all models considered. Model colors match those in the top panel of Figure~\ref{fig:temp-history}. The horizontal vertical red and blue lines indicate the $k$-modes corresponding to the Hubble flow across the entire simulation box and a single cell, respectively. Gray arrows mark the $k$-modes associated with comoving scales of 1 and 10 $h^{-1}$Mpc.
        At small $k$, we include observational limits from the DESI Collaboration \cite{DESI-DR1-FPS}, using $k_{\textrm{min}} = 0.0468\ \textrm{\AA}^{-1}$, converted to inverse velocity via $k = 0.0468\ \textrm{\AA}^{-1} \times \lambda_{\textrm{Ly}\alpha} (1 + z)/c$. At large $k$, the gray region shows the resolution limit from the high-resolution study of \citet{2019ApJ...872..101B}, corresponding to $\log_{10}(k_{\textrm{max}}/\textrm{s\ km}^{-1}) = -0.7$.
        The $k$-modes trace the velocity separation of absorbing \HI\ structures. The amplitude of the spectrum reflects the amount of neutral hydrogen available to absorb \Lya\ photons. As the Universe expands and reionization progresses, the overall power decreases due to the declining \HI\ fraction. Over the full velocity range probed, the results for all models lie nearly on top of one another. The corresponding relative differences between DDE and $\Lambda$CDM models are shown in Figure~\ref{fig:FPS-diff}.
    }
    \label{fig:FPS}
\end{figure*}

 \subsection{\Lya\ Transmitted Flux Power Spectrum} \label{subsec:Lya-FPS}
        
        The transmitted \Lya\ flux in a cell along a LOS skewer is computed using the standard relation $F = \exp(-\tau)$. To characterize flux fluctuations in the \Lya\ forest, we evaluate

\begin{equation}
\delta_F (u) = \frac{F(u) - \bar{F}}{\bar{F}}\,,
\end{equation}

\noindent where $\bar{F}$ denotes the average transmitted flux in a given snapshot, estimated as the median of the mean transmitted flux values across all skewers. In dimensionless form, the FPS is defined as

\begin{equation} \label{eq:FPS}
\Delta_F^2 (k) = \frac{1}{\pi} k P(k)\,,
\end{equation}

\noindent where the power spectrum $P(k)$ is computed as

\begin{equation} \label{eq:Pk}
P(k) = u_{\textrm{max}} \left\langle \left| \tilde{\delta}_F(k) \right|^2 \right\rangle
\end{equation}

\noindent for a velocity wavenumber $k = 2\pi / u$. The Fourier transform of the flux fluctuation field is given by

\begin{equation} \label{eq:delta_Fk}
\tilde{\delta}_F(k) = \frac{1}{u_{\textrm{max}}} \int_0^{u_{\textrm{max}}} e^{-iku} \delta_F (u) \textrm{d}u\,,
\end{equation}

\noindent where $u_{\textrm{max}} = H L / (1+z)$ is the Hubble velocity across the simulation box of comoving length $L$.

       The evolution of the dimensionless FPS during $4.0 < z < 1.317$ for the fiducial $\Lambda$CDM cosmology and the three $w_0 w_a$CDM models is shown in Figure~\ref{fig:FPS}. The range of scales spans from $k = 4 \times 10^{-4}$ km$^{-1}$ s to $k = 1.8$ km$^{-1}$ s. The peak amplitude decreases by more than an order of magnitude, from $\approx 2 \times 10^{-1}$ at $z = 4.0$ to $\approx 1.5 \times 10^{-2}$ at $z = 1.317$. As the comoving Hubble rate $H/(1+z)$ declines, the lowest $k$-mode shifts to larger physical scales. Over the redshift range shown, the amplitude at the lowest $k$ decreases by just over an order of magnitude.

	\begin{figure*}[t]
    \centering
    \includegraphics[scale=0.457]{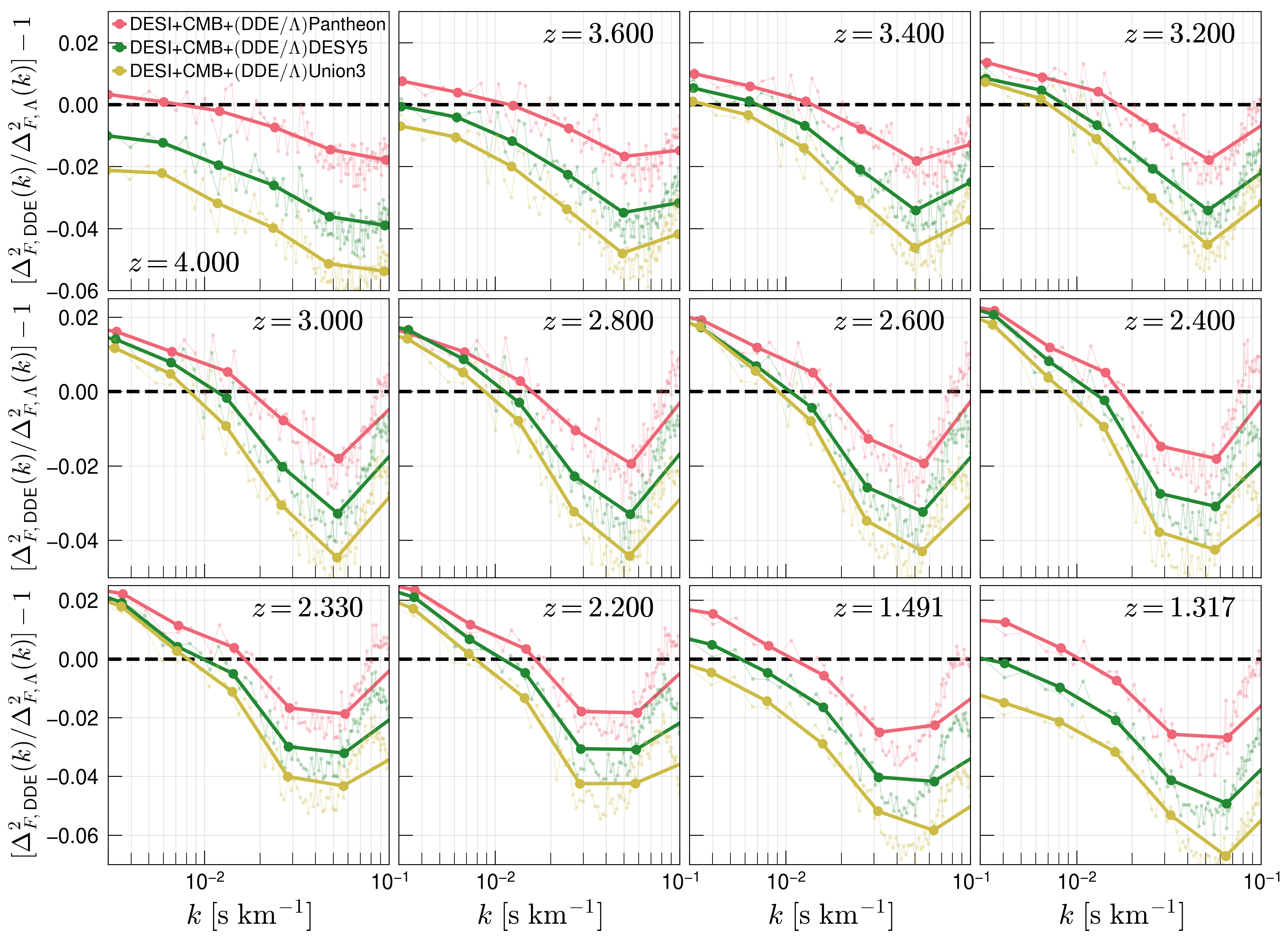}
  \caption{
    Evolution of the relative difference in the flux power spectrum. We show the comparison of a DDE and $\Lambda$CDM model for the Pantheon, DESY5, and Union3 cosmological models in peach, green, and yellow, respectively. At each scale $k$, the quantity $[\Delta_{F, \textrm{DDE}}^2(k) / \Delta_{F, \Lambda}^2(k)] - 1$ measures the fractional excess or deficit of power in a $w_0w_a$CDM cosmology relative to its $\Lambda$CDM counterpart. We average in bins of constant $\Delta \log_{10}(k/\textrm{s\ km}^{-1})$ and show the unbinned quantity with lower opacity. The difference is more pronounced at larger physical scales, indicating a scale-dependent feature, a spectral tilt, most clearly seen in the bottom row. For the DESI+CMB+Pantheon model, a slight excess of power appears at $k \lesssim 1.5 \times 10^{-2}\ \textrm{s\ km}^{-1}$ near $z = 3.2$.
}
    \label{fig:FPS-diff}
\end{figure*}
        
        \begin{figure*}[t]
    \centering
    \includegraphics[scale=0.365]{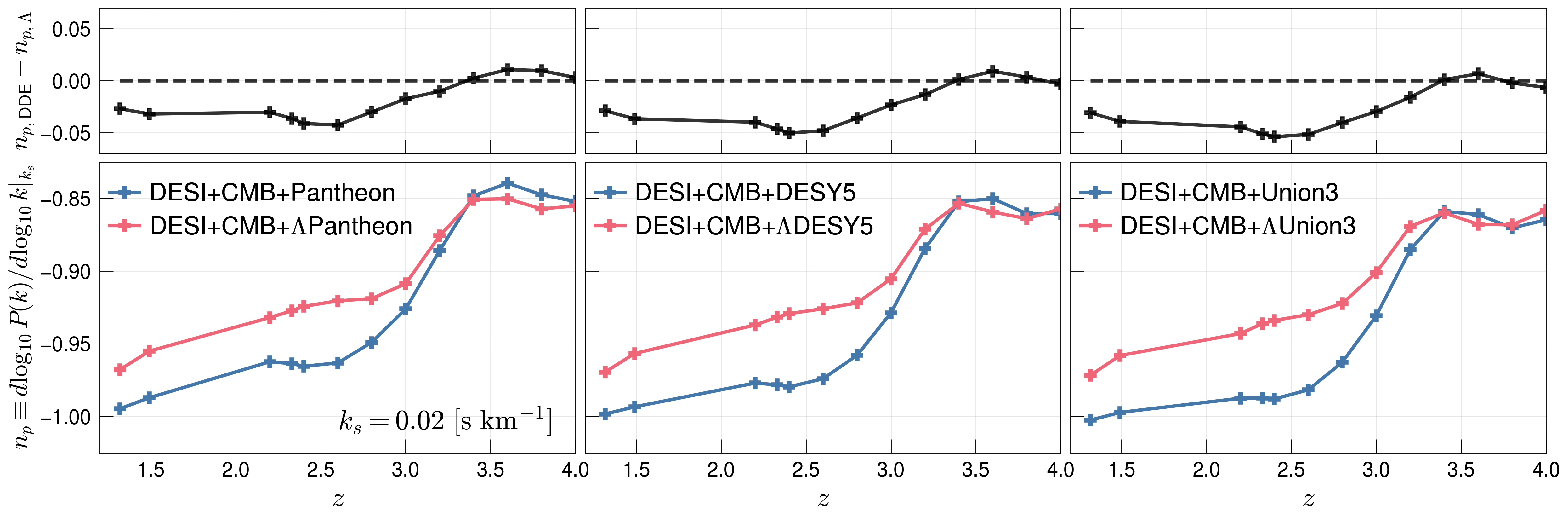}
    \caption{
    Evolution of the power spectrum slope $n_P$ evaluated at $k_s = 2 \times 10^{-2}$ s km$^{-1}$. From left to right, each panel shows the redshift evolution of $n_P$ for models based on the Pantheon, DESY5, and Union3 datasets. In blue, we show the $w_0w_a$CDM model; in peach, its $\Lambda$CDM counterpart. The top row displays the difference in slope between the two models. All calculations account for the effects of peculiar velocities. Following the completion of helium reionization at $z \approx 3$, the slope decreases more rapidly for DDE models.}
    \label{fig:FPS-slope}
\end{figure*}

        \subsection{Flux Power Spectrum Tilt} \label{subsec:Lya-FPS-tilt}

        As discussed in Section~\ref{subsec:sims-var}, we isolate the DE impact on the expansion history by comparing simulations with DDE against otherwise identical simulations with cosmological constant DE. We detail the impact of a DDE on the IGM by computing the differences of the dimensionless FPS, $\Delta^2(k) = kP(k)/\pi$, in a $w_0 w_a$CDM cosmology with its $\Lambda$CDM counterpart from Equation~\ref{eq:FPS}. Since we self-consistently model the non-equilibrium chemistry network for hydrogen, helium, and electrons and do not directly compare against observations, we do not match an effective optical depth when comparing between two cosmologies. The impact of matching the effective optical depth on the relative difference of the FPS is discussed in Appendix~\ref{appendix:FPS-taumatch}.
        
        Since we adopt the same comoving side lengths to our simulation boxes, the slightly different Hubble rate between a DDE and $\Lambda$CDM cosmology will land on different velocity bins. For the redshifts we analyze, we find $H_{\textrm{DDE}}(z) < H_{\Lambda}(z)$, so the \Lya\ spectra will extend slightly more in the $\Lambda$CDM cosmology compared to its DDE counterpart. Before taking the Fourier Transform of a DDE skewer, we mirror twice-over half of the number of cells required to reach the same velocity domain and ensure periodicity. We then take a linear interpolation to the same velocity bins as in the $\Lambda$CDM comparison cosmology. We show the relative FPS differences in Figure~\ref{fig:FPS-diff} after rebinning Equation~\ref{eq:Pk} to a set of constant $\Delta \log_{10}(k/\textrm{s\ km}^{-1})$ bins. In general, the magnitude and ordering of the differences in $\Delta^2(k)$ between DDE models and their $\Lambda$CDM counterparts follow the same pattern observed in the relative differences of the \Lya\ optical depth and IGM temperature: DESI+CMB+Pantheon, DESI+CMB+DESY5, and DESI+CMB+Union3. The first panel of Figure~\ref{fig:FPS-diff} shows that at $z = 4$, the primary difference between the $w_0 w_a$CDM cosmologies and their $\Lambda$CDM equivalents is an overall normalization shift across all $k$-modes.

        Over the redshift range $3.6 < z < 3.2$, we find a gradual relative increase in power at low-$k$ modes ($k/\textrm{s\ km}^{-1} < 2\times10^{-2}$) for each $w_0 w_a$CDM model. From $z = 4$ to $z = 3.2$, the FPS difference in the DESI+CMB+Union3 model steadily increases, approaching that of its $\Lambda$CDM counterpart at the lowest $k$-modes ($k < 7 \times 10^{-3}\ \textrm{s\ km}^{-1}$). Between $z=3.6$ and $z=3.2$, the relative power difference at $k = 5 \times 10^{-2}\ \textrm{s\ km}^{-1}$ remains approximately constant at $\leq 2\%$, $\approx 3.5\%$, and $\approx 4.5\%$ for the DESI+CMB+Pantheon, DESI+CMB+DESY5, and DESI+CMB+Union3 cosmologies, respectively. At the smallest scales displayed ($\log_{10} k/\textrm{s\ km} \approx -1$), we find the relative power increases by about 1\% from $z=3.6$ to $z=3.2$.
        
        At redshifts $z \leq 3.0$, we observe a clearer development of enhanced power at low $k$ and suppressed power at high $k$, resulting in a distinct spectral tilt, as shown in the bottom row. We expect deviations in the FPS from $\Lambda$CDM to be primarily driven by DDE after helium reionization completes ($z \approx 3$ for our adopted UV background), as the thermal evolution of the IGM becomes increasingly governed by the expansion history. For the DESI+CMB+Pantheon model, we identify a characteristic scale at $k \approx 1.8 \times 10^{-2}\ \textrm{s\ km}^{-1}$ where the FPS transitions from exhibiting excess to deficit power relative to its $\Lambda$CDM counterpart over the redshift range $z = 3.0$ to $z = 2.2$. Over the same interval, all three DDE models converge to similar relative FPS differences with respect to $\Lambda$CDM at the largest scales. During the subsequent phase of adiabatic expansion, from $z = 2.33$ to $z = 1.317$, the overall amplitude of the FPS difference changes, while the spectral tilt persists. Over this redshift range, the peak FPS amplitude at $k = 2 \times 10^{-2}\ \textrm{s\ km}^{-1}$ decreases from $4 \times 10^{-2}$ to $1.5 \times 10^{-2}$. Compared to their $\Lambda$CDM counterparts, the DDE models consistently exhibit more large-scale and less small-scale flux fluctuations down to $k \approx 3 \times 10^{-2}\ \textrm{s\ km}^{-1}$ .
        
        We compute the logarithmic derivative of the power spectrum to investigate the evolution of the spectral tilt. Assuming the power spectrum of Equation \ref{eq:Pk} evolves as a power law, we fit a linear polynomial for $\log_{10} P(k)$ as a function of $\log_{10} k$ in windows of 8 $k$-mode bins. We then fit a spline to the resulting slopes in each window to evaluate the expression
        
        \begin{equation}
            n_P\equiv\frac{\textrm{d} \log_{10} P(k) }{\textrm{d} \log_{10} k } \bigg\rvert_{k_s}
        \end{equation}
        
        \noindent at $k_{s} = 2 \times 10^{-2}\  \textrm{s\ km}^{-1}$ near the peak FPS amplitude, where the spectral tilt to transitions from positive to negative $\Delta^2_{F}$ values at $z=2.33$ between the DESI+CMB+Pantheon DDE and $\Lambda$CDM models. 
        
        Our results for the calculated $n_P$ values are presented in Figure~\ref{fig:FPS-slope}. The power spectrum computation includes the effects of peculiar velocities, capturing the contribution from the growth of small-scale structure. We find $n_P \approx -0.85$ relatively constant at redshifts $z > 3.4$, with DDE models displaying a slightly greater slope. For redshifts $z < 3.4$, we calculate a steeper logarithmic $P(k)$ slope for DDE when compared to $\Lambda$CDM models $n_{P, \textrm{DDE}} < n_{P, \Lambda}$. We find the spectral index difference between DDE and $\Lambda$CDM is most pronounced at redshifts $z \approx 2.5$ with $n_{P, \textrm{DDE}} - n_{P, \Lambda} \approx -0.05$. For all models considered, the difference decreases to $\approx -0.03$ at $z=1.317$ when $n_{P, \textrm{DDE}} \approx -1.0$ and $n_{P, \Lambda} \approx -0.97$. We notice that the overall shape of $n_P$ as a functtion of redshift for all three of our DDE models are similar, and likewise for our $\Lambda$CDM models.
    
    \begin{figure*}[t]
    \centering
    \includegraphics[scale=0.405]{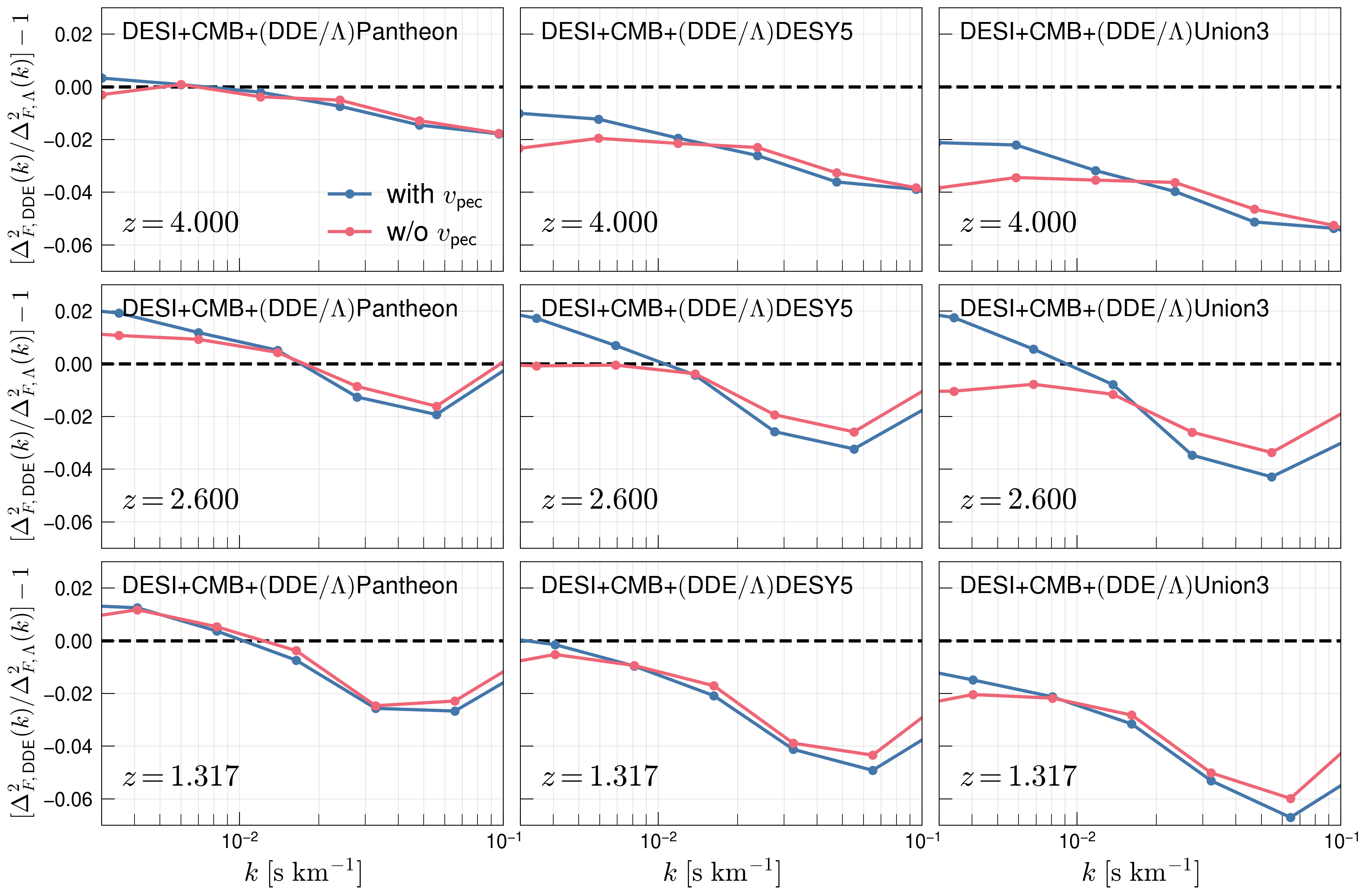}
    \caption{
        Comparison of the relative difference evolution of the FPS with and without peculiar velocities. From top to bottom, we show results at redshifts $z = 4$, $2.6$, and $1.317$. From left to right, panels correspond to the DESI+CMB+Pantheon, DESY5, and Union3 models. In peach (blue), we plot the relative FPS difference computed without (with) peculiar velocities. At all redshifts, excluding peculiar velocities reduces large-scale power ($k \leq 6 \times 10^{-3}$ s\ km$^{-1}$) in each model.
    }
    \label{fig:FPS-diff-vpec}
\end{figure*}

    \subsection{Impact of Peculiar Velocities on the Flux Power Spectrum}\label{subsec:Lya-FPS-pec}
                
   Peculiar velocities introduce an additional broadening effect by redistributing the optical depth of a cell across neighboring gas elements. Including peculiar velocities in the optical depth calculation suppresses small-scale power in the flux, reducing the amplitude at large $k$-modes. In addition, divergent velocity fields can raise the temperature of thermally pressurized gas regions. High-density regions, which grow according to the linear growth factor, are the origin of gravitationally induced peculiar velocities. The clustering of matter in redshift space is therefore altered, affecting the evolution of the large-scale linear matter power spectrum \citep{Kaiser}. While peculiar velocities damp small-scale flux fluctuations, the linear growth factor dominates the large-scale evolution, where structure formation proceeds in the linear regime.
   
	We compute the redshift evolution of the relative FPS difference with and without including $v_{\rm{pec}}$ for both the $\Lambda$CDM and DDE models, as shown in Figure~\ref{fig:FPS-diff-vpec}. At $z = 4$, removing peculiar velocities has minimal effect on the relative FPS difference for the DESI+CMB+Pantheon model. In contrast, the DESI+CMB+DESY5 and DESI+CMB+Union3 models exhibit a modest reduction in relative power difference at large scales ($k < 1.5 \times10^{-2}\ \textrm{s\ km}^{-1}$) when peculiar velocities are excluded. At small scales ($k > 6 \times 10^{-2}\ \textrm{s\ km}^{-1}$), all three models show slightly more power when excluding peculiar velocities.
	
	In the second row of Figure~\ref{fig:FPS-diff-vpec}, we find that removing peculiar velocities amplifies the large-scale relative FPS difference for the DESI+CMB+DESY5 and DESI+CMB+Union3 models in the range $k < 1.5 \times 10^{-2}\ \textrm{s\ km}^{-1}$ at $z = 2.6$. For $k < 8 \times 10^{-3}\ \textrm{s\ km}^{-1}$, the relative flux fluctuations are greater (lower) in the DDE models compared to their $\Lambda$CDM counterparts when including (excluding) peculiar velocities. In contrast, removing peculiar velocities in the DDE FPS calculation of DESI+CMB+Pantheon shows a slight decrease in the relative FPS difference at large scales, but is still greater than the $\Lambda$CDM counterpart. In the range $2 < k \times 10^{2}\ \textrm{km\ s}^{-1}$, the exclusion of peculiar velocities results in a slight increase in the relative FPS difference. Across all three models, we find that the exclusion of peculiar velocities decreases large-scale flux fluctuations, subsequently leading to a less distinct FPS spectral tilt.
    
	Our final redshift snapshot at $z = 1.317$ is shown in the third row of Figure~\ref{fig:FPS-diff-vpec}, exhibiting similar peculiar velocity effects to those observed at $z = 2.6$. For all models, we find a slightly lower large-scale relative FPS difference ($k < 6 \times 10^{-3}\ \textrm{s\ km}^{-1}$) when peculiar velocities are excluded. In the range $1 < k \times 10^{2}\ \textrm{km\ s}^{-1}$, removing peculiar velocities results in a slightly enhanced relative FPS difference across all models, while effectively landing on top of one another. The spectral tilt described in Section~\ref{subsec:Lya-FPS-tilt} remains present in all cases, and is more clearly visible in the relative FPS difference when peculiar velocities are included.
    
\section{Discussion}\label{sec:disc}

	We have shown that the IGM in a dynamical DE cosmology exhibits distinct properties compared to its counterpart with a cosmological constant. A brief discussion of our findings is presented here. In Section~\ref{subsec:DDE}, we place our results in the context of previous work and examine the impact of DDE on the FPS. Section~\ref{subsec:caveats} outlines some limitations in interpreting our results. Finally, in Section~\ref{subsec:implications}, we discuss the broader implications of our findings for DDE cosmologies.

	\subsection{Dynamical Dark Energy Impact on Flux Power Spectrum}\label{subsec:DDE}
	
	Our results differ from those of \cite{Viel2003} and \cite{Coughlin}, who also explored the impact of non-cosmological constant dark energy on the IGM and the \Lya\ forest before the DESI results. \citet{Viel2003} used semi-analytical models to compute the power spectrum (Equation~\ref{eq:Pk}) for $w$CDM cosmologies and reported a deviation from $\Lambda$CDM characterized by a constant normalization offset at $k < 3 \times 10^{-2}\ \textrm{s\ km}^{-1}$. At fixed $k$ in the redshift range $1.8 < z < 2.2$, all three of their models $[(w_0,w_a) = {(-0.4,0.0), (-0.6,0.0), (-0.8,0.0)}]$ produced higher $P(k)$ values than $\Lambda$CDM, with the most extreme case, $(w_0,w_a) = (-0.4,0.0)$, yielding nearly an order of magnitude excess. We find a qualitatively similar trend at $z = 2.2$, where the DESI+CMB+Pantheon and DESI+CMB+DESY5 models predict more large-scale power at $k < 10^{-2}\ \textrm{s\ km}^{-1}$ compared to their $\Lambda$CDM counterparts. For the DESI+CMB+Union3 model, a comparable effect occurs at slightly smaller scales, $k < 7 \times 10^{-3}\ \textrm{s\ km}^{-1}$. However, unlike \cite{Viel2003}, we find that the relative difference is not merely a normalization offset but exhibits a clear scale dependence.

    Using high-resolution dark matter–only $N$-body simulations, \cite{Coughlin} apply a semi-analytical prescription for gas based on density and temperature assignments to ``field" and ``halo" particles. They investigate four DDE models, $(w_0,w_a) = {(0.0,-3.0), (-2.0,0.0), (-2.0,2.0), (-1.1,1.3)}$, and report FPS deviations from $\Lambda$CDM. However, they emphasize that these discrepancies are not driven by DE and are subdominant to cosmic variance according to the Anderson-Darling statistic. Using identical sight-lines to compute the FPS, they find nearly indistinguishable power at large scales ($-2.5 < \log_{10}(k/\textrm{s\ km}^{-1}) < -2.0$), with only the model $(w_0,w_a) = (-1.1,1.3)$ showing a deficit in power. In contrast, we find a scale-dependent FPS difference between each $w_0w_a$CDM cosmology and its $\Lambda$CDM counterpart, indicating that the DDE  models explored here imprint a measurable deviation in the FPS.
    
    A recent study by \cite{DDEClustering} investigated the effect of DDE of the kind inferred by DESI on the matter power spectrum, clustering, and halo abundances using large, high resolution $N$-body simulations. They adopt cosmological parameters from \cite{Planck-Cosmo} as a reference model and compare against a model with the same parameters, changing only the DE equation of state to the $w_0 w_a$CDM DESI+CMB $(w_0,w_a)$ values from Table 3 of \cite{DESI-DR1-Cosmo}. At $z=2.03$ and $z=1.54$, they find DDE increases the power of dark matter fluctuations by $\approx 2\%$ across scales of $0.1 < k / (h \textrm{Mpc}^{-1} ) < 10$. At lower redshifts, they find a lower (greater) matter power spectrum amplitude at large-scales of $k < 2 h \textrm{Mpc}^{-1} $ (small-scales of $k > 2 h \textrm{Mpc}^{-1} $). Using hydrodynamical simulations, we find lower small-scale amplitude of $\Delta_F^2(k)$ in \Lya\ spectra. When comparing their reference $\Lambda$CDM model to the model with all $w_0 w_a$CDM DESI+CMB cosmological parameters, they find that the impact of varying other cosmological parameters, specifically highlighting the product $\Omega_{m,0} h^2$, dominate over the impact of DDE. When holding $\Omega_{m,0} h^2$ constant, \cite{DDEClustering} find the impact of DDE to suppress large-scale and enhance small-scale matter power fluctuations. To highlight the impact of DE on structure formation, further high-resolution simulations are warranted. 

	At late times ($z < 3.0$), we find a spectral tilt in the FPS that is qualitatively similar to that reported in warm dark matter models at higher redshift ($z > 4.0$). Previous studies have attributed the suppression of small-scale power to reduced density fluctuations, and the increase in large-scale power to an excess of close-to-mean density gas and lower mean transmitted flux caused by the free-streaming behavior of warm dark matter \citep{2024PhRvD.109d3511I, 2023PhRvD.108b3502V, 2013PhRvD..88d3502V, 2017MNRAS.471.4606A, 2016JCAP...08..012B, 2019MNRAS.489.3456G}. We also find less power at small scales of the FPS in the $w_0w_a$CDM models, as well as an increase in large-scale power, accompanied by a higher mean transmitted flux (i.e., a lower effective optical depth), with the distinguishing feature that the DDE-induced spectral tilt occurs at later times.
        
        We perform our calculations in redshift space but verify consistency when excluding peculiar velocities. Consistent with \cite{vpec_tau1, vepc_tau2}, we find that removing peculiar velocities yields a more optically thick IGM. At the same time, the percent-level relative differences in $\tau_{\rm eff}$ are amplified when peculiar velocities are excluded. Using radiative transfer simulations to study the effects of inhomogeneous reionization on the FPS, \cite{rad_vpec} report that higher gas temperatures induce stronger gradients in the peculiar velocity field of pressurized gas, leading to small-scale FPS suppression. The influence of peculiar velocities in shaping the FPS exceeds the free-streaming suppression from warm dark matter at scales $k > 10^{-1}\ \textrm{s\ km}^{-1}$ \citep{2023PhRvD.108b3502V}. Additionally, \cite{PRIYA-cosmo} find that a higher amplitude of the matter power spectrum at $k = 0.78\ \textrm{Mpc}^{-1}$ enhances large-scale FPS power and increases the peculiar velocity–induced suppression at small scales, particularly at $k > 2 \times 10^{-1}\ \textrm{s\ km}^{-1}$. In our analysis, excluding peculiar velocities reduces the relative FPS difference at large scales, while their inclusion tends to introduce greater variability in the relative FPS difference at small scales.

        Observations of quasars illuminating the \Lya\ forest along the line of sight provide a powerful probe of cosmic voids and serve as a sensitive testbed for cosmological physics and galaxy formation. High-resolution quasar spectra enable FPS measurements at small scales, as demonstrated by datasets such as KODIAQ \citep{2015AJ....150..111O}, SQUAD \citep{2019MNRAS.482.3458M}, XQ-100 \citep{2016A&A...594A..91L}, and archival Keck/HIRES and VLT/UVES data \citep{2019ApJ...872..101B, 2018ApJ...852...22W}. Medium-resolution spectra of large quasar samples provide robust constraints at larger scales, with surveys such as BOSS \citep{2013AJ....145...10D} and its extension eBOSS \citep{2016AJ....151...44D}. The DESI DR1 release further expands this capability with 700,000 medium-resolution \Lya\ quasar spectra \citep{2025JCAP...01..124A}.
	
	The measurement of the FPS using $\approx 500{,}000$ quasars from DESI DR1 is presented in \citep{DESI-DR1-FPS}. They report statistical uncertainties on the power spectrum (e.g., Equation~\ref{eq:Pk}) below 10\% across all $k$-modes and redshifts. Absolute statistical uncertainties reach the $10^{-2}$ level at all modes for $z < 3.4$. The total uncertainty budget is dominated by systematic effects, as shown in the bottom-right panel of Figure~11 in \citep{DESI-DR1-FPS}, where the ratio of systematic to statistical uncertainties is quantified. With future improvements aimed at reducing systematics and increasing quasar sample sizes in upcoming DESI data releases, we anticipate that precision measurements of the FPS will soon reach the sensitivity needed to detect DDE-induced deviations of the kind presented in this work—namely, relative differences of $+2\%$ at small $k$-modes and $-4\%$ at large $k$-modes in the DESI+CMB+Union3 cosmology.

\subsection{Caveats}\label{subsec:caveats}

Our results highlight several limitations that motivate future work aimed at more robustly isolating the effects of a DDE and performing joint cosmological parameter inference with observational data. The thermal state of the IGM is governed primarily by the balance between photoionization heating and adiabatic expansion cooling. The role of photoheating in DDE cosmologies could be more fully explored by marginalizing over the amplitude and redshift evolution of hydrogen and helium ionization (see Figure~4 of \citep{2022ApJ...933...59V} for the effect of ionization history on IGM temperature).

A potential limitation of our study is the assumption of a spatially uniform ionizing background. Realistic modeling of reionization, particularly the patchy and quasar-driven nature of HeII reionization, is essential to accurately assess the imprint of DDE on the FPS. Simulations that incorporate spatially inhomogeneous ionization fields using radiative transfer have shown increased small-scale FPS power at high redshift ($z > 5$) for $k \leq 0.03\ \textrm{s\ km}^{-1}$ \citep{10.1093/mnras/stz2807, 10.1093/mnras/sty968, 10.1093/mnras/stac3761}. At lower redshifts ($2 < z < 4$), \citep{10.1093/mnras/stz1388} find that ionization inhomogeneities may induce measurable FPS differences of $3.3$–$6.5\%$ at $z = 4$ and $0.35$–$0.75\%$ at $z = 2$. Other studies implementing radiative transfer coupled with hydrodynamics to model inhomogeneous HeII reionization also report enhanced small-scale FPS power resulting from altered thermal histories \citep{La_Plante_2017, 10.1093/mnras/staa1850}. In the context of \Lya-emitting galaxy clustering, \cite{rad_vpec_LAE} show that radiative diffusion of the \Lya\ line can have a larger effect than peculiar velocities. In addition, we do not account for gas cooling via metal-line emission, assuming instead a composition of pristine hydrogen and helium. This simplification, together with the omission of AGN feedback, may introduce a bias at the few percent level in FPS-based cosmological parameter constraints \citep{10.1093/mnras/staa1242}.

	In the DDE cosmological models considered in this study, the rate of expansion that governs adiabatic cooling is determined by the energy-density fractions, the present-day Hubble parameter $H_0$, and the dark energy parameters $(w_0, w_a)$. Varying $H_0$ alters the physically inferred value of $\Omega_b h^2$, thereby changing the total amount of hydrogen and helium. While we adopt three specific $(w_0, w_a)$ pairs, a more comprehensive analysis would sample the posterior distribution presented in Figure~11 of \citep{DESI-DR2-Cosmo}. Varying the matter energy-density could modify both the expansion history and the growth of structure, which are expected to leave imprints on the FPS.

Because the \Lya\ forest is sensitive to small-scale structure, a more complete treatment would marginalize over the effects of linear matter power spectrum parameters. Combining thousands of quasar spectra with hydrodynamic simulations, \citep{2006MNRAS.365..231V} demonstrated that the FPS can constrain $\Omega_m$, $n_s$, and $\sigma_8$, establishing its value as a cosmological probe of structure formation. As noted in Section~\ref{subsec:DDE}, changes to structure formation induced by warm or fuzzy dark matter have also been studied using hydrodynamic simulations. Recent work has employed emulators trained on high-resolution simulations to efficiently reproduce \Lya\ forest observables and constrain the amplitude and spectral index of the matter power spectrum at small scales \citep{PRIYA, 2023ApJ...944..223P}. While such emulators capture the impact of structure growth efficiently, they typically do not resolve the full complexity of IGM physics. Since the combined effects of a DDE and variations in structure growth parameters (e.g., matter density, amplitude, and spectral slope of the power spectrum) on the IGM remain largely unexplored, high-resolution hydrodynamic simulations are essential to accurately quantify how DDE influences the diffuse baryons probed by the \Lya\ forest.

    \subsection{Implications for Dynamical Dark Energy}\label{subsec:implications}
    
    The phenomenological $w_0-w_a$ parameterization of the DE equation of state considered in this study describes a wide range of DE models after mapping the observationally constrained Hubble rate $H(z)$ and distances $D(z)$ to a time evolving equation of state $w(z)$ \citep{linder_dynamics_2008, de_Putter_2008}. 
    The models considered for this study all cross the ``phantom divide'' [ $w(z) \leq -1$] some time in the recent past ($z \approx 0.4$), which violates the null energy condition ($\rho + P \geq 0$) for simple physical models of DE. However, there are other DE models with
    similar expansion histories and Hubble rates that do not violate the null energy condition.
    For instance, \cite{Shlivko_2024} show that different classes of simple, physically-motivated models can satisfy the null energy condition, while occupying an equivalent $w_0-w_a$ parameter space that leads to a crossing of the phantom line. Since a physical interpretation of the $w_0-w_a$ model can sometimes prove challenging, \cite{Shajib_2025} advocate for a thawing ($w_a < 0$) scalar field that recently ($z \approx 2$) begins to exponentially grow from $w=-1$ in the early Universe. The model in \cite{Shajib_2025} has only a single parameter $w_0$ to fit, which can be interpreted as the scalar field mass or the initial value of the scalar field. In principle, either of these DE models, or similar classes of models that change the expansion histories or Hubble rates relative to $\Lambda$CDM, could leave observable imprints in the \Lya\ forest FPS by changing the adiabatic cooling and IGM temperature history. We note that these models would also change the linear growth function for the development of cosmic structure, which may also have an effect on the  forest. We plan to simulate such models in future work.

\section{Conclusions}

   We study the impact of DDE in $w_0w_a$CDM cosmologies by analyzing its effects on the thermal history of the IGM and the structure of the \Lya\ forest using a focused suite of hydrodynamic simulations. Holding $\sigma_8$, $n_s$, and $\Omega_b h^2$ fixed, we run simulations with three distinct sets of cosmological parameters ($\Omega_{m,0}$ and $H_0$) for both $w_0w_a$CDM and $\Lambda$CDM cosmologies. The DDE parameter choices are motivated by constraints from DESI DR1. Our main findings are summarized below:

	\begin{itemize}
    \item We quantify the effect of DDE on the Hubble rate, cosmic age, and linear growth factor relative to matched $\Lambda$CDM models. The Hubble rate in all DDE cosmologies follows a common trend, leading to an older Universe compared to $\Lambda$CDM down to $z \approx 0.25$. Each DDE model exhibits a transition from suppressed to enhanced linear growth at a characteristic redshift, depending on the underlying $(w_0, w_a)$ parameters.
    
    \item We calculate and compare the effective optical depth, low-density temperature, and FPS for the DDE and $\Lambda$CDM cosmologies considered. Relative differences between these quantities are typically at the few percent level.
    
    \item The simulations show a slightly warmer low-density IGM in DDE cosmologies, with relative temperature differences in the range $0.1\% < [T_{\textrm{LD,DDE}} / T_{\textrm{LD},\Lambda}] - 1 < 2\%$ for $z < 4$, assuming the same cosmic UV background evolution.
    
    \item The effective optical depths in DDE models are lower by 1–4\% relative to their $\Lambda$CDM counterparts over the redshift range $z \approx 1.5$–4. The medians of the logarithmic distributions of the \HI\ fraction and neutral hydrogen density differ by up to $-0.5\%$.
    
    \item We find that DDE models introduce a spectral tilt in the transmitted flux power spectrum relative to $\Lambda$CDM, characterized by more relative power at large scales and less at small scales. This scale-dependent feature is most prominent at $z < 3$ and cannot be accounted for by rescaling the effective optical depth or excluding peculiar velocities, indicating that DDE leaves a distinct physical imprint on the structure of the \Lya\ forest.
    
\end{itemize}

    Our results point to an intriguing possibility of a DDE imprint on \Lya\ measurements. The complex relationship between the matter power spectrum, meta-galactic UV background, and potential baryonic influences from galactic feedback on the FPS necessitates high-resolution hydrodynamic simulations to distill the impact of a DDE on the \Lya\ forest.

\begin{acknowledgments}
    This research used resources of the Oak Ridge Leadership Computing Facility at the Oak Ridge National Laboratory, which is supported by the Office of Science of the U.S. Department of Energy under Contract No. DE-AC05-00OR22725. An award of computer time was provided by the INCITE program. 
    This material is based upon work supported by the National Science Foundation Graduate Research Fellowship Program under Grant No Award 2240310. Any opinions, findings, and conclusions or recommendations expressed in this material are those of the author(s) and do not necessarily reflect the views of the National Science Foundation.
    We acknowledge use of the lux supercomputer at UC Santa Cruz, funded by NSF MRI grant AST 1828315. This research was supported by NASA Astrophysics Theory Program Grant 80NSSC22K0814.

\end{acknowledgments}

\appendix

\section{Lookback Time}

	An alternative method of describing the expansion history of a desired cosmology is with the dimensionless lookback time ($H_0t$), as a function of redshift by solving the equation

        \begin{equation} \label{eqn:lookback}
            H_0t(z) = \int_0^{z} \frac{\textrm{d}z'}{(1+z') \xi(z')}  \,.
        \end{equation}
	
\begin{figure}
            \centering
            \includegraphics[scale=0.448]{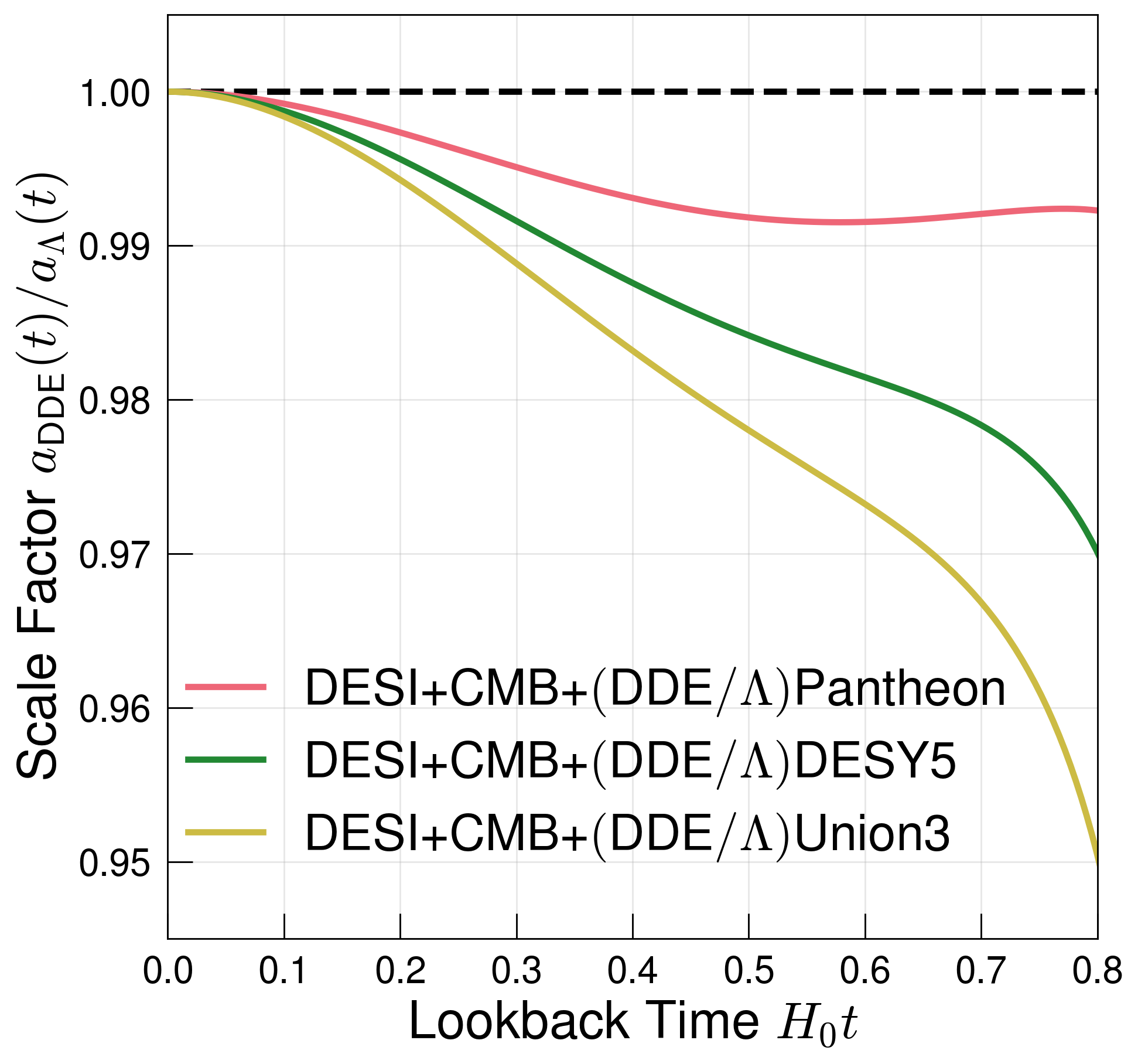}
            \caption{The expansion history for $w_0w_a$CDM cosmologies, $a_{\sf{DDE}}(t)$, normalized to their $\Lambda$CDM counterparts, $a_{\Lambda}(t)$, as a function of dimensionless lookback time $t$. The same colors as in Figure \ref{fig:cosmo-info}. The peach line shows the expansion history for the DESI+CMB+Pantheon divided by the expansion history for the DESI+CMB+$\Lambda$Pantheon cosmology, as described in Table \ref{tab:table1}. The case is the same for the green line showing the impact of replacing the DE equation of state in the DESI+CMB+DESY5 model with $w=-1$. The yellow line shows the expansion history of the DESI+CMB+Union3 model normalized by the same cosmology with a $(w_0,w_a)=(-1,0)$ parameter set.}
            \label{fig:expansion-history}
        \end{figure}

	After solving for the lookback time at a given redshift for each cosmology using Equation \ref{eqn:lookback}, we describe the scale factor $a=(1+z)^{-1}$ (i.e., expansion history) as a function of the lookback time. Our representation of the DDE parameters is shown in Figure \ref{fig:expansion-history}. The least to most discrepant expansion history of a DDE cosmological model with respect to their $\Lambda$CDM counterpart follows from Figure 6 of \cite{DESI-DR1-Cosmo}: Pantheon, DESY5, and Union3. All $w_0w_a$CDM cosmologies predict a slightly smaller scale factor at fixed lookback time, with larger discrepancies at increasing lookback time.

        At a lookback time of $H_0t=0.5$, the relative expansion history difference for the DESI+CMB+Pantheon model reaches $\approx$1\% and stays roughly the same at larger lookback times. For the same lookback time, the scale factor is predicted to be 1.8\% and 2.2\% smaller for the DESI+CMB+DESY5 and DESI+CMB+Union3 $w_0 w_a$CDM models compared to their $\Lambda$CDM cosmologies, respectively. Looking back to 80\% that of a Hubble time $(1/H_0)$, we find that length scales in a $w_0 w_a$CDM cosmology reach only 95\%, 97\%, and 99\% that of the $\Lambda$CDM versions for the DESI+CMB+Pantheon, DESI+CMB+DESY5, and DESI+CMB+Union3 models respectively. We expect the \HI\ density to be reduced at fixed lookback time for a given $\Omega_{b,0}$, a trend that should manifest in baryonic properties of the IGM.

\section{Optical Depth Calculation}\label{app:opt-depth}

    We measure the optical depth along a skewer by setting $u_0$ as the cell-centered Hubble flow velocity. The integral contribution from the $i$-th cell to the optical depth at the $j$-th cell in Equation \ref{eq:tau_u0} becomes

    \begin{equation}
         \int_{u_{i - 1/2}}^{u_{i + 1/2}} \frac{1}{v_{\textrm{th}, i} \pi^{1/2}}   \exp\left[-\left(\frac{u_i - u_j}{v_{\textrm{th}, i}} \right)^2 \right] \textrm{d}u_i \,.
    \end{equation}

    We complete a substitution of variables $\mu = (u_{i} - u_j) / v_{\textrm{th}, i}$ with limits $\mu_\pm = (u_{i\pm 1/2} - u_j) / v_{\textrm{th}, i}$ to integrate the exponential. We now have

    \begin{equation}
        \frac{1}{\pi^{1/2}}
         \int_{\mu_{-}}^{\mu_{+}}  \exp\left[- \mu^2\right] \textrm{d}\mu \,,
    \end{equation}

    \noindent where we use the error function to get

    \begin{equation} \label{eq:errf-1}
        \frac{1}{2} \left( \erf\left[ \frac{u_{i+1/2} - u_j}{v_{\textrm{th}, i}} \right] - \erf\left[ \frac{u_{i- 1/2} - u_j}{v_{\textrm{th}, i}} \right] \right) \,.
    \end{equation}

    We note that the physical velocity of a cell $u_i$ is the summation of the peculiar velocities $v_i$ and the cell-centered Hubble flow velocity $v_{H,i}$. The boundary values $u_{i+1/2}$ are found by adding half of the Hubble flow across one cell, $\Delta v_H = u_{\textrm{max}} / N_{\textrm{LOS}}$ (where $u_{\textrm{max}}$ is defined after Equation \ref{eq:delta_Fk}, and $N_{\textrm{LOS}}$ is the number of LOS cells), to the physical velocities $u_{i\pm1/2} = u_i \pm \Delta v_H / 2$. We rearrange the error function argument to place the evaluation around the $j$-th cell to find

    \begin{equation}
        u_{i\pm1/2} - u_j = u_i - u_{j, \mp 1/2} \,.
    \end{equation}

    Using the property that the error function is an odd function, we rewrite Equation \ref{eq:errf-1} as

    \begin{equation} \label{eq:errf-2}
        \frac{1}{2} \left( \erf\left[ \frac{u_{j+1/2} - u_i}{v_{\textrm{th}, i}} \right] - \erf\left[ \frac{u_{j- 1/2} - u_i}{v_{\textrm{th}, i}} \right] \right) \,.
    \end{equation}
    
    \begin{figure*} 
            \centering
            \includegraphics[scale=0.457]{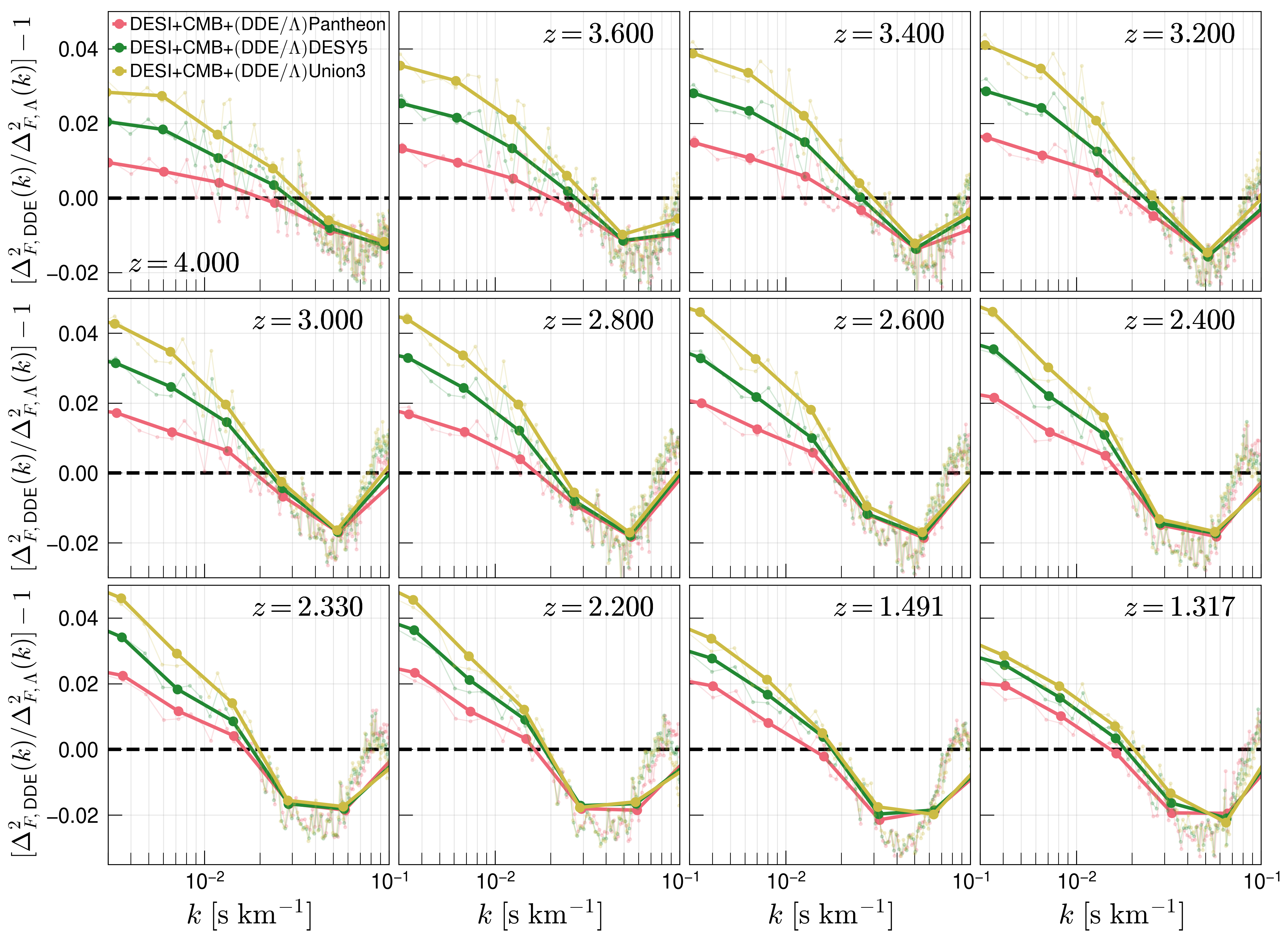}
            \caption{Relative difference in the flux power spectrum, after artificially boosting the \HI\ density for a DDE model to match the effective optical depth of its $\Lambda$CDM counterpart. Same colors as in Figure \ref{fig:cosmo-info}. We bin to values of constant $\Delta \log_{10}(k/\textrm{s\ km}^{-1})$, with the unbinned quantity shown with low opacity. As in Figure \ref{fig:FPS-diff}, we find a relative difference between a DDE and its $\Lambda$CDM counterpart that is dependent on the spectral scale and redshift.}
            \label{fig:FPS-diff-taumatch}
    \end{figure*}

    \noindent We complete the measurement by looping over each $j$-th cell. We note that the measurements are completed at the center of the cell, so the peculiar velocity of the cell taking the measurement is zero, $v_j = 0$, and the only contribution to $u_{j \pm 1/2}$ is the Hubble flow along the left or right interface. Using the notation after Equation \ref{eq:opt-depth-discrete}, $u_{j+1/2} = v_{\textrm{R},H,j}$ and $u_{j-1/2} = v_{\textrm{L},H,j}$.

\section{Impact of Matching \Lya\ Effective Optical Depth on the Flux Power Spectrum} \label{appendix:FPS-taumatch}

    The goal of matching the effective optical depth is to marginalize over the unknown UV background field and fix the opacity at a specific length scale. The UV background adopted in this study was designed to fine tune the model from \cite{10.1093/mnras/stz222} by changing the amplitude and timing of hydrogen and helium reionization in a $\Lambda$CDM model with cosmological parameters from \cite{Planck-Cosmo}. As mentioned in Section \ref{subsec:caveats}, an expanded study could vary the UV background field in order to jointly fit the impact of DDE with FPS observations. To illuminate whether our primary result of the spectral tilt arises from an overall effective optical depth normalization (or $\sigma_8$ normalization), we artificially boost the \HI\ density in $w_0w_a$CDM cosmologies to match the effective optical depth in each of their $\Lambda$CDM model counterparts. We show the relative measurement for all three alternative cosmologies in Figure \ref{fig:FPS-diff-taumatch}. 
    
    A primary result of this study, the FPS shows scale- and redshift-dependent differences between DDE and corresponding $\Lambda$CDM models, remains readily apparent. The overall impact of matching the effective optical depth manifests as an amplitude change at all $k$-mode scales, with no impact on the spectral tilt shown in Figure \ref{fig:FPS-diff}. At large scales ($\log_{10} (k\ \textrm{km\ s}^{-1}) \leq -2$), we calculate a larger DDE FPS amplitude compared to a $\Lambda$CDM model. Reproducing the effect of the spectral tilt requires a more complicated set of cosmological parameters than simply increasing or decreasing $\sigma_8$. By matching to the effective optical depth of the $\Lambda$CDM model, the physical effects that induce differences between the DDE and cosmological constant model FPS may be obfuscated. We therefore present our results without matching the optical depths, while noting that, without marginalizing over the effects of the UV background on the FPS normalization, the FPS amplitude cannot be independently calibrated.

\bibliography{codes_cite, IGM, DarkEnergy, DESIrefs, LymanAlphaFPSCosmo, LymanAlphaObservations, Supernova}

\end{document}